\documentclass[
  aps,
  pra,
  reprint,
  superscriptaddress,
  amsmath,
  amssymb,
  floatfix,
  bibnotes, %
  longbibliography
]{revtex4-2}
\usepackage[T1]{fontenc}
\usepackage[american]{babel}
\usepackage{graphicx} %
\usepackage{url}
\usepackage{makecell} %
\usepackage[dvipsnames]{xcolor}
\usepackage{booktabs}
\usepackage{capt-of}
\usepackage{placeins}
\usepackage{acro}
\acsetup{
  first-style = long-short,
  subsequent-style = short,
  single = true,
  single-style = long
}

\DeclareAcronym{mzi}{
  short = MZI,
  long = Mach--Zehnder interferometer
}
\DeclareAcronym{spdc}{
  short = SPDC,
  long = spontaneous parametric down-conversion
}
\DeclareAcronym{pbs}{
  short = PBS,
  long = polarizing beam splitter
}
\DeclareAcronym{hwp}{
  short = HWP,
  long = half-wave plate
}
\DeclareAcronym{bs}{
  short = BS,
  long = beam splitter
}
\DeclareAcronym{lp}{
  short = LP,
  long = linear polarizer
}
\DeclareAcronym{spad}{
  short = SPAD,
  long = single-photon avalanche diode
}
\DeclareAcronym{cc}{
  short = CC,
  long = coincidence counter
}
\DeclareAcronym{ttl}{
  short = TTL,
  long = transistor-transistor logic
}

\DeclareAcronym{igm}{
  short = IGM,
  long = intergalactic medium
}
\DeclareAcronym{lambdacdm}{
  short = \ensuremath{\Lambda}\textrm{CDM},
  long = \ensuremath{\Lambda} cold dark matter
}
\DeclareAcronym{ir}{
  short = IR,
  long = infrared
}
\DeclareAcronym{uvfs}{
  short = UVFS,
  long = ultraviolet fused silica
}

\DeclareAcronym{aeronet}{
  short = AERONET,
  long = Aerosol Robotic Network
}
\DeclareAcronym{aod}{
  short = AOD,
  long = aerosol optical depth
}
\DeclareAcronym{tpw}{
  short = TPW,
  long = total precipitable water
}
\DeclareAcronym{metar}{
  short = METAR,
  long = Meteorological Aerodrome Report
}
\DeclareAcronym{goes18}{
  short = GOES-18,
  long = Geostationary Operational Environmental Satellite 18
}
\DeclareAcronym{cod}{
  short = COD,
  long = cloud optical depth
}
\DeclareAcronym{irsa}{
  short = IRSA,
  long = Infrared Science Archive
}
\DeclareAcronym{utc}{
  short = UTC,
  long = Coordinated Universal Time
}
\DeclareAcronym{ra}{
  short = RA,
  long = right ascension
}
\DeclareAcronym{dec}{
  short = Dec,
  long = declination
}
\DeclareAcronym{dssii}{
  short = DSS-II,
  long = second-generation Digitized Sky Survey
}

\DeclareAcronym{nasa}{
  short = NASA,
  long = National Aeronautics and Space Administration
}
\DeclareAcronym{noaa}{
  short = NOAA,
  long = National Oceanic and Atmospheric Administration
}
\DeclareAcronym{ucsb}{
  short = UCSB,
  long = {University of California, Santa Barbara}
}
\DeclareAcronym{nsf}{
  short = NSF,
  long = National Science Foundation
}
\DeclareAcronym{nrt}{
  short = NRT,
  long = National Science Foundation Research Traineeship
}

\usepackage[protrusion=true,expansion=true]{microtype}

\newcommand{\runinheading}[1]{%
  \par\medskip\noindent\emph{#1}\textemdash\nobreak\ignorespaces
}
\usepackage[hidelinks]{hyperref}
\MakeRobust{\%}

\begin{document}
\raggedbottom

\title{Complementarity Test with Unmeasured, Permanently Inaccessible Path Markers}
\author{Paul Gauthier}
\email{paul@paulg.com}
\affiliation{Independent Researcher, Santa Barbara, California, USA}

\author{Sahil Patel}
\affiliation{Department of Electrical and Computer Engineering,
  University of California, Santa Barbara, California 93106, USA}

\author{Sean Doan}
\affiliation{Department of Physics,
  University of California, Santa Barbara, California 93106, USA}

\author{Galan Moody}
\affiliation{Department of Electrical and Computer Engineering,
  University of California, Santa Barbara, California 93106, USA}

\date{September 21, 2026}
\hypersetup{
  pdftitle={Complementarity Test with Unmeasured, Permanently Inaccessible Path Markers},
  pdfauthor={Paul Gauthier; Sahil Patel; Sean Doan; Galan Moody}
}
\begin{abstract}
  In prior complementarity tests, path markers were measured and remained causally
  accessible, leaving the collapse-locality loophole open.
  We address this loophole using a quantum eraser that isolates the encoding
  of path information in the joint quantum state from its later measurement
  and causal accessibility.
  We launch idler photons---the sole path markers for entangled signal photons
  in an interferometer---on outgoing null trajectories while the joint state
  remains in a coherent superposition of both interferometer path alternatives.
  Our flat \(\Lambda\)CDM transmission model predicts that
  \(73\)--\(82\%\)
  of launched idlers will propagate forever without absorption, scattering,
  or environmental path-record formation.
  For these modeled survivors, each path marker remains perpetually unmeasured,
  and propagation along its outgoing trajectory precludes
  later causal contact with its interferometric record.
  Unconditioned signal detections show no statistically significant
  launch-induced change in interference, as standard quantum theory predicts.
  Normalized to the modeled survivors, the 95\% confidence upper bound on any
  launch-induced fringe is \(0.16\) of full
  restoration.
\end{abstract}
\maketitle

\section{Introduction}
\label{sec:introduction}

Complementarity is often stated in terms of whether path information is
available in principle. Zeilinger gives a canonical formulation for the
double-slit experiment:
\begin{quote}
  ``It is sufficient to destroy the
  interference pattern, if the path information is accessible in
  principle from the experiment or even if it is dispersed in the
  environment and beyond any technical possibility to be recovered, but
  in principle still `out there.'
  The absence of any such information is
  \textit{the essential criterion} for quantum interference to appear.''%
  ~\cite{Zeilinger1999Experiment}
\end{quote}

Operationally, this criterion encompasses three distinct elements that have
coincided in previous complementarity tests. \textit{Encoding} writes path
information into the joint quantum state by entangling path alternatives with
distinguishable marker states. \textit{Measurement} forms a record of the path
information in a detector, absorber, environment, or observer. Whether, when,
and where a measurement produces a definite outcome is precisely the
measurement problem. Given this ambiguity, \textit{causal accessibility}
concerns whether measurement of the path marker can causally influence the
formation of the interference record.

\begin{figure}[!t]
  \centering
  \includegraphics{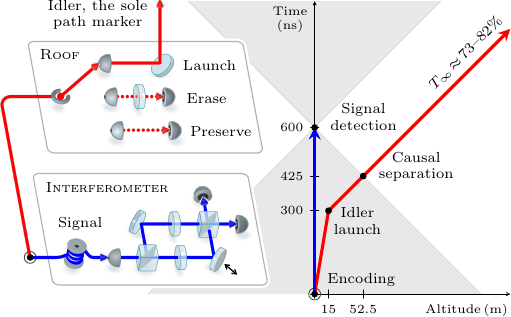}
  \caption{
    Experimental schematic and causal geometry.
    Each polarization-entangled idler photon carries the only path marker for its
    signal in the interferometer.
    At the roof, a common path supports launching the idlers,
    projecting them to erase the marker, or detecting them with the marker
    preserved; three paths are drawn for clarity.
    Pair creation encodes path information at~\(0\,\mathrm{ns}\).
    The idler is launched on an outgoing null
    trajectory at \(\sim300\,\mathrm{ns}\) and exits the past light cone
    of the signal-detection event at \(\sim425\,\mathrm{ns}\).
    The signal is detected at \(\sim600\,\mathrm{ns}\).
    Our transmission model
    of the launch optics, atmosphere, Milky Way, and \acl*{igm}
    predicts that
    \(T_\infty\approx73\)--\(82\%\)
    of idlers will propagate unmeasured forever.
  }
  \label{fig:cones}
\end{figure}

Previous complementarity experiments did not operationally isolate these
elements. Across many timing and separation regimes, they showed that
interference does not depend on the causal ordering of the presumed measurement
events~\cite{ScullyDruhl1982,Kim2000DelayedChoiceEraser,Ma_2016}. Even so, the
relevant path markers were always measured, and causal separation between the
marker and interference measurement sites was only fleeting. Even in the
\(144\,\mathrm{km}\) La Palma--Tenerife experiment, causal influence could pass
between the measurement sites after only~\(0.5\,\mathrm{ms}\)~\cite{Ma_2013}.

Kent showed why such fleeting separation may be insufficient by noting that a
detector click need not be the true measurement
event~\cite{Kent2005CausalQuantumTheory,Kent_2020}. In delayed-collapse
theories, definite record formation may occur later. These actual
outcome-forming events may lie in a common causal region even when the nominal
clicks were spacelike separated. Prior tests may therefore have established
causal separation only at the onset of the measurement process. This is the
essential form of the collapse-locality loophole.

We report what is, to our knowledge, the first complementarity test to separate
path-marker encoding from subsequent measurement and causal
accessibility~(Fig.~\ref{fig:cones}). We launch polarization-entangled idler
photons on outgoing null trajectories, carrying the sole path markers for their
twin signals in a \ac{mzi}. In this causal geometry, each idler exits the past
light cone of its signal-detection event while the joint state remains in a
coherent superposition of both interferometer path alternatives.

Our transmission model of the launch optics, atmosphere, Milky Way, and
\ac{igm} predicts that \(73\)--\(82\%\) of launched idlers propagate
indefinitely without absorption, scattering, or environmental record formation.
For these idlers, the causal separation is permanent, not fleeting. The path
marker never localizes in any detector or environment, and no laboratory record
can overtake a photon receding on a null trajectory. No future event can gain
causal access to both the path marker and the corresponding interferometric
record.

\begin{figure}[t]
  \centering
  \includegraphics{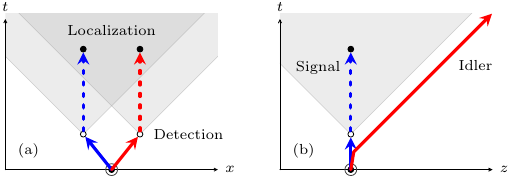}
  \caption{
    (a) The essential collapse-locality loophole.
    Although the detector clicks are spacelike separated, later record-forming
    events can lie in the causal future of both clicks.
    In complementarity tests,
    the resulting unconditioned signal statistics may therefore not represent a trace over an
    unmeasured coherent marker.
    (b)~Under the launch condition, the signal-detection
    and localization events have no causal access to surviving idler clicks
    or localized records.
  }
  \label{fig:loophole}
\end{figure}

Figure~\ref{fig:loophole} contrasts the launch geometry with the
collapse-locality loophole that may have remained open in prior tests. Standard
quantum theory equates these physically distinct configurations through the
partial-trace identity
\[
  \rho_s
  =
  \underbrace{
    \operatorname{Tr}_i \rho_{si}
    \vphantom{\sum_k p_k \rho_{s|k}}
  }_{
    \substack{
      \mathstrut\text{Idler unmeasured,}\\
      \mathstrut\text{inaccessible}
    }
  }
  =
  \underbrace{\sum_k p_k \rho_{s|k}\mathpunct{.}}_{
    \substack{
      \mathstrut\text{Idler measured,}\\
      \mathstrut\text{outcome } k \text{ ignored}
    }
  }
\]
The identity predicts our null result immediately; indeed, any launch-induced
  fringe could allow signaling. But no-signaling is a consequence of the
  identity, not an independent guarantee of it. Because prior tests measured
  their path markers, if collapse is sufficiently delayed, their unconditioned
  statistics realized only the outcome-marginalization side. The launch
  condition realizes the untested partial-trace side. For the modeled majority
  of idlers, no measurement outcome \(k\) ever occurs, and the partial trace
  describes the physical situation rather than serving as bookkeeping.

The experiment is therefore a loophole-closing null test of a central quantum
equivalence. As in later loophole-closing Bell
tests~\cite{Salart_2008,Hensen2015LoopholeFree,Giustina2015LoopholeFree,Shalm2015LoopholeFree,BIG_Bell_Test_2018},
its significance lies not in a surprising outcome but in eliminating reliance
on a physical assumption in a foundational test. Cosmic Bell tests carried this
logic to cosmological scales, pushing the causal origin of measurement settings
into the deep past~\cite{Rauch2018CosmicBell}; our launch geometry pushes the
causal fate of the path marker into the infinite future.

To perform the test, we first use quantum-eraser controls to verify that
coherent path markers reach the launch section of the apparatus. We then launch
the path-marked idlers and search for a launch-induced interference fringe in
unconditioned signal detections. A transmission model estimates the rate of
path markers predicted to survive indefinitely, and any such fringe is
normalized to that rate.

Across 6.6 hours of integration and \(10^8\) signal detections, we observed no
statistically significant launch-induced change in unconditioned signal
interference, consistent with quantum theory. After normalization to the
modeled ensemble of indefinitely surviving idlers, the 95\% confidence upper
bound on any launch-induced fringe amplitude is \(0.16\) of full restoration.

\section{Surviving-idler ensembles}
\label{transmission}
\label{supp-transmission}

Formally, single-photon interference in a two-path interferometer is set by the
overlap of the idler marker states ($|m_{\mathrm{U}}\rangle_i$ and
$|m_{\mathrm{L}}\rangle_i$) entangled with the signal's upper and lower path
alternatives in the interferometer ($|\mathrm{U}\rangle_s$ and
$|\mathrm{L}\rangle_s$, respectively). For a balanced state, the visibility of
unconditioned signal interference (the signal-singles visibility $\mathcal V$)
obeys complementarity
\begin{equation}
  |\Psi\rangle =
  \frac{|\mathrm{U}\rangle_s|m_{\mathrm{U}}\rangle_i+
    e^{i\phi}|\mathrm{L}\rangle_s|m_{\mathrm{L}}\rangle_i}{\sqrt2},
  \;
  \mathcal V=
  |\langle m_{\mathrm{U}}|m_{\mathrm{L}}\rangle|.
  \label{eq:marker-overlap}
\end{equation}
Thus, when the two idler marker states are orthogonal, they contain,
in principle, complete which-path information about the signal photon and suppress
unconditioned signal interference. Quantum theory predicts this holds whether the idler
is later detected, measured in an erasing basis, or decoheres into the
environment.

We define a surviving idler as one modeled to avoid these outcomes. Survivors
are neither absorbed nor scattered, and form no environmental record correlated
with the path marker. Coherent propagation effects, including redshift,
diffraction, and free-space beam expansion, do not constitute loss provided
they create no such record. The relevant quantity is the overlap of the two
propagated marker states in Eq.~\eqref{eq:marker-overlap}, not their fidelity
to the launched state.

We use two nested reference ensembles of launched idlers. The first consists of
idlers modeled to survive unmeasured forever; no future event can access both
these path markers and their interferometric outcomes. The broader ensemble
requires survival only through the Milky Way. Even an interaction immediately
after Galactic transit could not influence the laboratory for at least
\(10^4\)\,years, far beyond any delayed-localization interval. This ensemble
therefore addresses the essential collapse-locality loophole using only a
finite-path transmission model.

Table~\ref{tab:transmission} provides the modeled survival fractions of the two
ensembles, defined as
\[
  T_{\mathrm{fin}}
  \equiv T_{\mathrm{opt}}T_{\mathrm{atm}}T_{\mathrm{MW}},
  \qquad
  T_{\infty}
  \equiv T_{\mathrm{fin}}T_{\mathrm{IGM},\infty}.
\]

\begin{table}[t]
  \caption{Modeled transmission factors and reference-ensemble survival fractions for idlers.
    Ranges span the experimental datasets. Displayed values are rounded;
    products use full precision.
  }
  \label{tab:transmission}
  \centering
  \begin{tabular*}{\columnwidth}{@{}l@{\extracolsep{\fill}}rr@{}}
    \toprule
    Component or ensemble & Transmission & Loss \\
    \midrule
    Launch optics, \(T_{\mathrm{opt}}\) & \(0.99\) & \(1.5\%\) \\
    Atmosphere, \(T_{\mathrm{atm}}\) & \(0.86\)--\(0.92\) & \(8.5\)--\(14\%\) \\
    Milky Way, \(T_{\mathrm{MW}}\) & \(0.84\)--\(0.94\) & \(5.6\)--\(16\%\) \\
    \Acl*{igm}, \(T_{\mathrm{IGM},\infty}\) & \(0.98\) & \(1.6\%\) \\
    \midrule
    Finite-path ensemble, \(T_{\mathrm{fin}}\) & \(0.75\)--\(0.83\) & \(17\)--\(25\%\) \\
    Infinite-future ensemble, \(T_{\infty}\) & \(0.73\)--\(0.82\) & \(18\)--\(27\%\) \\
    \bottomrule
  \end{tabular*}
\end{table}

The launch optics comprised a mirror and a silica window, whose reflectance and
transmission, respectively, set \(T_{\mathrm{opt}}\). Atmospheric transmissions
were computed with clear-sky, zenith, direct-beam libRadtran
calculations~\cite{Emde2016libRadtran,Mayer2005libRadtran,Gasteiger2014REPTRAN}
constrained by \acp{metar}~\cite{AviationWeatherMETAR_20260305_08} and either
satellite observations~\cite{GOES18AOD,GOES18COD,GOES18TPW} or lunar
\ac{aeronet} observations~\cite{Holben1998,Giles2019,Schafer2026LunarAERONET}.
Milky Way transmissions came from NASA/IPAC’s Galactic DUST Reddening and
Extinction service~\cite{irsa_dust,Schlegel_1998,Schlafly_2011} along each
launch zenith.

Only the infinite-future ensemble depends on propagation through the \ac{igm}.
Dust extinction and galaxy interception give comparable contributions to its
modeled attenuation, with a smaller contribution from Thomson scattering. We
integrate all three along the idler's future null trajectory to the
cosmological limit \(a\to\infty\). In a flat \ac{lambdacdm} cosmology with
Planck 2018 parameters~\cite{planck:2018},
\[
  \tau_{\mathrm{IGM},\infty}
  =
  \frac{c}{H_0}
  \int_1^\infty
  \frac{k_{810,0}g(a)+n_{\mathrm{gal},0}\pi r_{\mathrm{gal}}^2
    +n_{e,0}\sigma_T}
  {a^4\sqrt{\Omega_m a^{-3}+\Omega_\Lambda}}\,da .
\]
The present-day dust opacity \(k_{810,0}\) is normalized to inferred mean
  intergalactic extinction~\cite{Menard2010IGMDust}, with \(g(a)\) accounting
  for wavelength-dependent attenuation as the idler redshifts. We model
  galaxies as opaque disks with present-day number density
  \(n_{\mathrm{gal},0}\) and effective radius
  \(r_{\mathrm{gal}}\)~\cite{Craig1996}. Thomson scattering uses the
  present-day mean free-electron density \(n_{e,0}\) and cross section
  \(\sigma_T\). Cosmological dilution at late times ensures convergence of the
  integral.

Appendix~\ref{app:transmission-details}\ gives full-precision inputs and the
complete model, including negligible attenuation channels.
Appendix~\ref{transmission-robustness}\ tests the conclusion's sensitivity to
the transmission model rather than assigning uncertainties to individual
factors.

\section{Experiment}
\label{sec:experimental-implementation}

\subsection{Apparatus}

\begin{figure*}[t]
  \centering
  \includegraphics{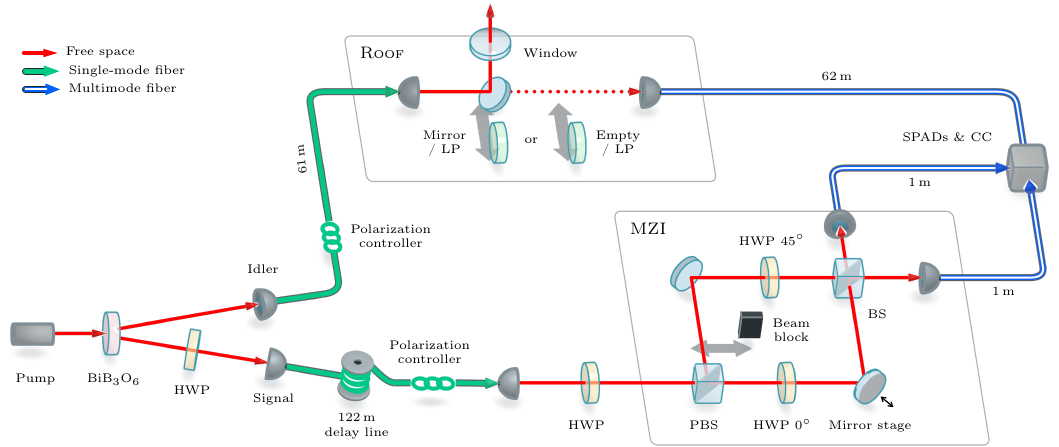}%
  \caption{
    A crossed-crystal \acl*{spdc} source
    based on bismuth borate (BiB\(_3\)O\(_6\)) emitted
    polarization-entangled signal--idler pairs in the Bell state
    \(\lvert\Phi^+\rangle =
    (\lvert\mathrm{H}\rangle_s\lvert\mathrm{H}\rangle_i
    + \lvert\mathrm{V}\rangle_s\lvert\mathrm{V}\rangle_i)/\sqrt{2}\).
    Signals traveled through a \(122\,\mathrm{m}\) delay line to a
    polarizing \acf*{mzi},
    where a \acf*{pbs} routed \(\mathrm H_s\) and \(\mathrm V_s\) into separate arms.
    \Acfp*{hwp} in the arms ensured that the signals
    reached the final \acf*{bs}
    horizontally polarized,
    carrying no path record.
    The idlers traveled through a \(61\,\mathrm{m}\) fiber to a free-space
    section on the roof.
    In the launch--erase run, a flip mount alternated between a launch mirror
    and a \acf*{lp} that erased the marker.
    In the preserve--erase run, it alternated between an empty slot that
    preserved the marker and the \acs*{lp}.
    Signals in all conditions and idlers in the erase and preserve conditions
    traveled through multimode fibers to \acfp*{spad}, whose outputs were
    recorded by a \acf*{cc}.
    The two upstream signal \acsp*{hwp} and the beam block were used for
    polarization calibration.
  }
  \label{fig:setup}
\end{figure*}

We generated \(810\,\mathrm{nm}\) polarization-entangled signal--idler pairs in
the Bell state $|\Phi^+\rangle = (|\mathrm{H}\rangle_s|\mathrm{H}\rangle_i
+|\mathrm{V}\rangle_s|\mathrm{V}\rangle_i)/\sqrt2$ using a crossed-crystal
BiB\(_3\)O\(_6\) \ac{spdc} source~\cite{Kwiat1999Ultrabright} pumped by a
continuous-wave \(405\,\mathrm{nm}\) laser. Figure~\ref{fig:setup} shows these
components and the full apparatus used for data collection.

The signal remained in the ground-floor laboratory, traversed a
\(122\,\mathrm{m}\) fiber delay, and entered a polarizing \ac{mzi}. Its input
\ac{pbs} routed \(\mathrm H_s\) and \(\mathrm V_s\) into separate arms, where
\acp{hwp} rotated both polarizations to horizontal before recombination. Thus
only idler polarization marked the path, as established by the erase--preserve
coincidence-fringe comparison defined in
Sec.~\ref{sec:normalization-conventions}.

The idler traveled through \(61\,\mathrm{m}\) of fiber to a free-space section
on the building roof, where the launch, erase, and preserve conditions were
implemented. The preserve condition sent the idler directly to a \ac{spad} with
its \(\mathrm H/\mathrm V\) marker intact, while the erase condition projected
it onto a state complementary to \(\mathrm H/\mathrm V\) before detection. In
the launch condition, a \(45^\circ\) mirror sent the idler upward through a
silica window and into the open atmosphere.

The same optical train supported the launch, erase, and preserve conditions.
This common-path design reduced alignment systematics, while movable elements
selected the relevant roof operations for each run. During the launch--erase
run, the roof flip mount alternated between the launch mirror and an erase
assembly containing the polarizer and a baffle that blocked ambient light
entering through the sky-facing silica window. During the preserve--erase run,
the window was capped, and the flip mount alternated between the erase \ac{lp}
and an empty slot, which left the idler path marker intact.

Appendix~\ref{sec:experimental-setup} describes the source and detection
system, acquisition protocol, and data corrections.

\subsection{Causal geometry}
\label{sec:causal-geometry}
\label{main-app:causal-geometry}

\label{app:causal-geometry}%
When the launch mirror was inserted, the experiment causally separated the
path-marked idler from its partner signal before the signal was detected at
the \ac{mzi} exits. Figure~\ref{fig:cones} summarizes the timing geometry. We take
\(t=0\) at pair creation and \(z=0\) at the ground-floor laboratory. The idler
reached the launch mirror after \(61\,\mathrm{m}\) of fiber at
\((t_{\mathrm{launch}},z_{\mathrm{launch}})\approx
(300\,\mathrm{ns},15\,\mathrm{m})\), while the signal traversed
\(122\,\mathrm{m}\) of delay fiber and reached a detector at
\(t_{\mathrm{det}}\approx600\,\mathrm{ns}\).

After launch, the idler followed a vertical null trajectory,
\[
  z_i(t)\approx z_{\mathrm{launch}}+c(t-t_{\mathrm{launch}}).
\]
Its intersection with the signal-detection past light cone satisfies
\[
  z_i(t_{\mathrm{exit}})=c(t_{\mathrm{det}}-t_{\mathrm{exit}}),
\]
giving \((t_{\mathrm{exit}},z_{\mathrm{exit}})\approx
  (425\,\mathrm{ns},52.5\,\mathrm{m})\). The marker therefore left that past
  light cone \(125\,\mathrm{ns}\) after launch and \(175\,\mathrm{ns}\) before
  detection. At detection it was already about \(105\,\mathrm{m}\) above the
  laboratory and thereafter followed
\[
  z_i(t)\approx c(t-t_{\mathrm{det}})+105\,\mathrm{m},
\]
parallel to the \ac{mzi} future-light-cone boundary. No carrier of the
  laboratory record can overtake an infinite-future survivor.

The launch--erase flip mount was set seconds before each acquisition, in the
common causal past of every recorded event. We therefore do not claim a
spacelike-separated setting choice. This timing fixes the experiment's causal
scope. We target Kent's essential form of the collapse-locality loophole, which
he identifies as the more strongly motivated version~\cite{Kent_2020}. The
extended form additionally permits local hidden variables to depend on the
measurement setting independently of any collapse outcome and would require a
spacelike-separated launch--erase choice.

A launch-correlated signal-singles fringe would violate the standard
partial-trace identity. Such a fringe would not, in this implementation alone,
demonstrate superluminal signaling. A signaling protocol would require fast,
spacelike-separated switching, which this apparatus did not implement. Its
causal objective was to separate the marker from the formation of the signal's
interferometric record and from that event's causal future.

\subsection{Data collection}
\label{sec:acquisition}

Six datasets, labeled D1--D6, were collected March 5--8, 2026, under favorable
atmospheric conditions and with the dust-rich Galactic plane away from zenith.
Each dataset comprised two consecutive runs, one alternating launch and erase
and the other alternating preserve and erase. Each run contained nine\ \ac{mzi}
phase scans. Each scan used 11 piezo-voltage settings spanning about 1.5 fringe
periods, with \(10\,\mathrm{s}\) acquisitions at each setting. At each \ac{mzi}
phase setting, the two active conditions were measured before the piezo
advanced, using a precision flip mount to toggle between them.

\section{Analysis}
\label{sec:data-analysis}

\subsection{Interference-fringe fits}
\label{scan-fits}
\label{supp-scan-fits}

For each condition, the two \ac{mzi} output rates were fit jointly across the
nine\ scans as functions of the piezo mirror-stage voltage \(x\), with separate
exit-specific mean rates and opposite fringe signs,
\[
  R_{1,2}(x)
  =
  \bar R_{1,2}
  \left[
    1\pm \mathcal V\cos\left(2\pi(x-x_{0,k})/P\right)
    \right].
\]
Each scan had a phase offset \(x_{0,k}\) to model slow interferometer drift.
  The median of the \(x_{0,k}\) values defines the nominal phase of each
  condition’s joint fit.

The period was first left free in a fit to the high-visibility erase
coincidences in the launch--erase run. The period was then fixed within each
dataset for the other condition fits and the paired launch--erase contrast fit.
We report the fitted-period uncertainty as a \(1\sigma\) statistical
uncertainty, as are all other quoted uncertainties unless noted otherwise.

The derived exit-specific modulation amplitudes are \(\Delta R_i=\mathcal V\bar
R_i\). Their sum gives the total two-output fringe amplitude, whose propagated
variance from the final covariance matrix of the simultaneous two-output fit is
\[
  \begin{aligned}
    A & =\Delta R_1+\Delta R_2=\mathcal V(\bar R_1+\bar R_2), \\
    \sigma_A^2
      & =\sigma_{\Delta R_1}^2+\sigma_{\Delta R_2}^2
    +2\operatorname{Cov}(\Delta R_1,\Delta R_2).
  \end{aligned}
\]
The covariance term includes the dependence induced by the shared visibility;
  the marginal exit uncertainties are not treated as independent measurements.
  The separately fitted erase and preserve conditions have no modeled
  cross-condition covariance, so
\[
  \sigma^2(R_{\mathrm{pm}})
  =
  \sigma_A^2(\mathrm E)+\sigma_A^2(\mathrm P).
\]

Standalone and joint signal-singles fringe fits used Poisson count likelihoods,
whereas coincidence and paired contrast fits used weighted least squares with
propagated counting variances. To account for excess scatter, we multiplied the
covariance matrices from the least-squares fits by the reduced chi-square
whenever it exceeded one.

Per-scan fit results for all six datasets are included in the archived data and
analysis code~\cite{repo2026}.

\subsection{Paired launch--erase contrast fit}
\label{null-bound}

For each dataset, we tested for a launch-only singles fringe of arbitrary phase
while allowing for a residual fringe common to launch and erase.

We jointly fit the interleaved records using the dark-subtracted output
contrast \(z=R_1-R_2\). Each launch scan was paired by index with the
corresponding erase scan from the same run, retaining each record's measured
piezo mirror-stage voltage in the fit. For point \(j\) in scan \(k\), we
modeled the contrasts as
\[
  \begin{aligned}
    z_{{\mathrm E},k,j}
    ={} & a_{{\mathrm E},k}                 \\
        & +C_0\cos\theta_{{\mathrm E},k,j}  \\
        & +S_0\sin\theta_{{\mathrm E},k,j},
  \end{aligned}
\]
and
\[
  \begin{aligned}
    z_{{\mathrm L},k,j}
    ={} & a_{{\mathrm L},k}                                    \\
        & +(C_0+C_{\mathrm{add}})\cos\theta_{{\mathrm L},k,j}  \\
        & +(S_0+S_{\mathrm{add}})\sin\theta_{{\mathrm L},k,j}.
  \end{aligned}
\]
Here \(\theta_{c,k,j}=2\pi(x_{c,k,j}-x_{0,k})/P\) for \(c\in\{\mathrm
  E,\mathrm L\}\), where \(\mathrm E\) and \(\mathrm L\) denote erase and
  launch. The erase coincidences fixed the phase reference. The parameters
  \(a_{{\mathrm E},k}\) and \(a_{{\mathrm L},k}\) absorb scan-wise offsets,
  \((C_0,S_0)\) describe the common residual fringe, and
  \((C_{\mathrm{add}},S_{\mathrm{add}})\) describe a launch-only fringe.

Let \(\mathbf q=(C_{\mathrm{add}},S_{\mathrm{add}})\), and let \(\hat{\mathbf
q}\) denote its best-fit estimate. The added-fringe amplitude is
\(A_{\mathrm{add}}=\|\mathbf q\|\), and its fitted value is \(\hat
A_{\mathrm{add}}=\|\hat{\mathbf q}\|\). For an arbitrary added phase, \(\mathbf
q\) was bounded using the 95\% confidence ellipse
\[
  \mathcal{E}_{0.95}
  =
  \left\{
  \mathbf q\,\middle|\,
  \left(\mathbf q-\hat{\mathbf q}\right)^{\mathrm T}
  \Sigma^{-1}
  \left(\mathbf q-\hat{\mathbf q}\right)
  \le \chi^2_{2,0.95}
  \right\}.
\]
Here \(\chi^2_{2,0.95}\) is the 95th percentile of the chi-square
  distribution with two degrees of freedom. The resulting amplitude bound is
\[
  A_{\mathrm{add},95}
  =
  \max_{\mathbf q\in\mathcal{E}_{0.95}}\|\mathbf q\|.
\]

\subsection{Surviving-idler normalization}
\label{sec:normalization-conventions}

\begin{figure*}[t]
  \centering
  \includegraphics{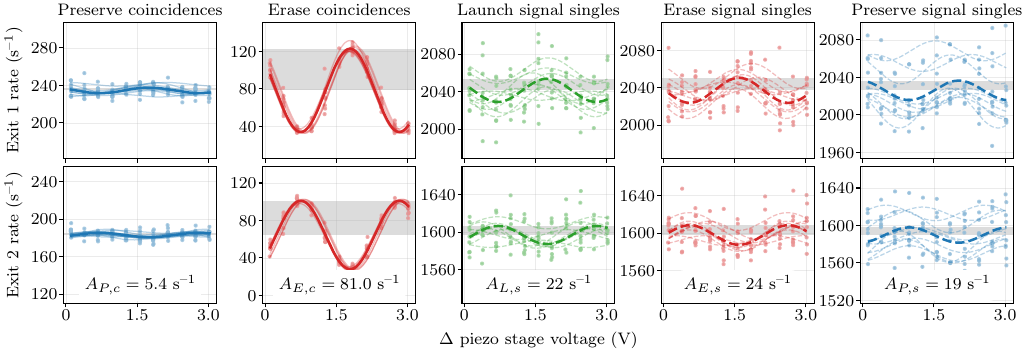}%
  \caption{
    Interference scans for
    the preserve, erase, and launch conditions
    from the illustrative dataset D5.
    With the idler path marker intact, the preserve condition gave a total
    two-exit coincidence fringe amplitude of
    \(A_{\mathrm{P},\mathrm{c}}=(5.4\pm1.0)\,\mathrm{s}^{-1}\).
    The erase condition gave an amplitude of
    \(A_{\mathrm{E},\mathrm{c}}=(81.0\pm1.0)\,\mathrm{s}^{-1}\).
    The excess,
    \(R_{\mathrm{pm}}
    \equiv A_{\mathrm{E},\mathrm{c}}-A_{\mathrm{P},\mathrm{c}}
    =(75.6\pm1.4)\,\mathrm{s}^{-1}\),
    establishes a conservative rate of coherent path-marked idlers reaching the roof,
    and
    \(R_\infty
    \equiv T_\infty R_{\mathrm{pm}}
    \approx 56\,\mathrm{s}^{-1}\)
    estimates the rate expected to survive unmeasured forever.
    The launch-condition signal singles nevertheless retained the controls'
    low residual fringe amplitude.
    Common vertical scales and alignment facilitate comparison of the shaded
    fringe amplitudes.
    Each panel contains 9\ scans over 11 settings, with
    \(10\,\mathrm{s}\) integrations; faint points and curves show individual
    scans, and thick curves show joint fits.
  }
  \label{fig:results}
\end{figure*}

Idlers detected in the preserve condition retained their path markers, whereas
projecting them onto the erasing state recovered coincidence interference. We
denote the total coincidence-fringe amplitudes across the two \ac{mzi} exits in
the erase and preserve conditions by \(A_{\mathrm{E},\mathrm{c}}\) and
\(A_{\mathrm{P},\mathrm{c}}\), respectively. The excess recovered by erasure
provides a conservative lower bound on the coherent path-marked rate reaching
the roof launch section. We therefore define
\[
  R_{\mathrm{pm}}
  \equiv A_{\mathrm{E},\mathrm{c}}-A_{\mathrm{P},\mathrm{c}}.
\]

The infinite-future transmission \(T_\infty\) is the modeled fraction of idlers
that survive forever while remaining unmeasured and permanently causally
inaccessible. Their coherent path-marked rate is
\[
  R_\infty
  \equiv T_\infty R_{\mathrm{pm}}.
\]
Analogously, the finite-path transmission \(T_{\mathrm{fin}}\) is the modeled
  fraction of idlers that survive unmeasured through the Milky Way, giving
\[
  R_{\mathrm{fin}}
  \equiv T_{\mathrm{fin}} R_{\mathrm{pm}}.
\]
These rates set the fringe-amplitude scales for full restoration in the
  infinite-future and finite-path analyses, respectively. The corresponding
  dimensionless restoration fractions are
\[
  \eta_\infty
  \equiv \frac{A_{\mathrm{add}}}{R_\infty},
  \qquad
  \eta_{\mathrm{fin}}
  \equiv \frac{A_{\mathrm{add}}}{R_{\mathrm{fin}}},
\]
where \(A_{\mathrm{add}}=\|\mathbf q\|\). For either fraction, zero denotes
  no restored fringe and unity denotes full restoration.

Full restoration places \(\mathbf q\) on the circle \(\|\mathbf q\|=R_\infty\)
at an unknown phase. Its Euclidean radial gap from the best-fit amplitude is
\(R_\infty-\hat A_{\mathrm{add}}\). The exclusion significance uses the full
covariance of \((C_{\mathrm{add}},S_{\mathrm{add}})\), including unequal
uncertainties and their correlation. Define \(\|\mathbf v\|_{\Sigma^{-1}}
\equiv(\mathbf v^{\mathrm T}\Sigma^{-1}\mathbf v)^{1/2}\) and \(\mathbf
u(\phi)=(\cos\phi,\sin\phi)\). Then
\[
  Z
  =
  \min_{\phi}
  \left\|R_\infty\mathbf u(\phi)-\hat{\mathbf q}\right\|_{\Sigma^{-1}}.
\]
For the finite-path exclusion, the same definition uses \(R_{\mathrm{fin}}\)
  in place of \(R_\infty\). Full restoration is a one-parameter circle within
  the unrestricted two-quadrature model, so after profiling over the phase,
  \(Z^2\) asymptotically follows a chi-square distribution with one degree of
  freedom. \(Z\) is thus the nominal Gaussian-equivalent significance. The
  normalized estimates, limits, and \(Z\) values are conditional on the stated
  transmissions and fitted \(R_{\mathrm{pm}}\) values. Because the transmission
  factors lack probabilistic uncertainties, we report nominal significances
  exceeding \(5\sigma\) simply as \(>5\sigma\) in the text; tables retain raw
  \(Z\) values as diagnostics.

\subsection{Combined analysis}

\begin{table*}[t]
  \newcommand{\DatasetAnalysisTableStyle}{%
    \small\setlength{\tabcolsep}{6pt}%
  }
  \caption{
    Dataset-level rates and bounds, ordered chronologically.
    \(R_{\mathrm{pm}}\) is the coincidence-derived path-marked rate, and
    \(A_{\mathrm{add},95}\) is the 95\% upper bound on the launch-only fringe
    amplitude; both are independent of the transmission model.
    For each ensemble, \(R\) is the transmission-scaled path-marked rate,
    \(\eta_{95}\) is the 95\% upper bound on the restoration fraction, and
    \(Z\) is the nominal Gaussian-equivalent significance for excluding full restoration.
    Rates are in \(\mathrm{s}^{-1}\), and \(R_{\mathrm{pm}}\) uncertainties are
    marginal \(1\sigma\). The normalized bounds and \(Z\) values are
    conditional on the stated transmission model.
  }
  \label{tab:dataset-analysis}
  \centering
  \begingroup
  \providecommand{\DatasetAnalysisTableStyle}{%
    \footnotesize\setlength{\tabcolsep}{1.5pt}%
  }%
  \DatasetAnalysisTableStyle{}
  \begin{tabular*}{\linewidth}{@{}l@{\extracolsep{\fill}}cccccccc@{}}
    \toprule
    & \multicolumn{2}{c}{Model independent}
    & \multicolumn{3}{c}{Infinite future}
    & \multicolumn{3}{c}{Finite path} \\
    \cmidrule(lr){2-3}\cmidrule(lr){4-6}\cmidrule(l){7-9}
    Dataset
    & $R_{\mathrm{pm}}$
    & $A_{\mathrm{add},95}$
    & $R_\infty$
    & $\eta_{\infty,95}$
    & $Z_\infty$
    & $R_{\mathrm{fin}}$
    & $\eta_{\mathrm{fin},95}$
    & $Z_{\mathrm{fin}}$ \\
    \midrule
    D1 & $61.3\pm1.4$ & $15$ & $46$ & $0.33$ & $9.0$ & $47$ & $0.32$ & $9.2$ \\
    D2 & $61.7\pm1.3$ & $14$ & $45$ & $0.29$ & $9.3$ & $46$ & $0.29$ & $9.5$ \\
    D3 & $70.4\pm1.3$ & $17$ & $58$ & $0.30$ & $12$ & $58$ & $0.29$ & $12$ \\
    D4 & $49.0\pm1.2$ & $12$ & $39$ & $0.31$ & $9.2$ & $39$ & $0.31$ & $9.4$ \\
    D5 & $75.6\pm1.4$ & $19$ & $56$ & $0.34$ & $10$ & $56$ & $0.34$ & $10$ \\
    D6 & $63.4\pm1.6$ & $18$ & $49$ & $0.37$ & $9.0$ & $49$ & $0.36$ & $9.2$ \\
    \bottomrule
  \end{tabular*}
  \endgroup
\end{table*}

For each dataset \(d=1,\ldots,6\), the paired fit gave a launch-only quadrature
estimate \(\widehat{\mathbf q}_d\) with covariance \(\Sigma_d\). With
\(R_{\infty}^{(d)}=T_{\infty}^{(d)}R_{{\rm pm}}^{(d)}\), we modeled
\begin{equation*}
  \begin{gathered}
    \widehat{\mathbf q}_d
    \sim\mathcal N\!\left[
      \eta R_{\infty}^{(d)}\mathbf u(\phi_d),\,\Sigma_d
      \right],\\
    \mathbf u(\phi)=(\cos\phi,\sin\phi),
  \end{gathered}
\end{equation*}
using a common nonnegative restoration fraction \(\eta\) and an independently
profiled phase \(\phi_d\) for each dataset. Because \(\eta=0\) is a boundary
point and the phases are unidentified there, parametric simulations calibrated
the null likelihood-ratio test. The one-sided profile-likelihood endpoint used
the nominal cutoff. Separate simulations checked its coverage over the
prespecified restoration-fraction grid and phase configurations and did not
require a larger, simulation-derived cutoff.

The finite-path analysis used the same likelihood with
\(R_{\mathrm{fin}}^{(d)}\) in place of \(R_\infty^{(d)}\).

\section{Results}
\label{sec:results}

We begin with D5 to illustrate the analysis applied to each dataset. The excess
total coincidence amplitude recovered in the erase condition relative to the
preserve condition establishes a conservative lower bound on the coherent
path-marked rate reaching the roof launch section. In the preserve condition,
idlers were detected with their path markers intact, and the total coincidence
fringe amplitude across the two \ac{mzi} exits was
\(A_{\mathrm{P},\mathrm{c}}=(5.4\pm1.0)\,\mathrm{s}^{-1}\), with visibility
\(0.013\pm0.002\). In the launch--erase run, projecting the idlers onto the
erasing state recovered coincidence interference fringes with amplitude
\(A_{\mathrm{E},\mathrm{c}}=(81.0\pm1.0)\,\mathrm{s}^{-1}\) and visibility
\(0.567\pm0.006\). The excess gives
\[
  R_{\mathrm{pm}}
  \equiv
  A_{\mathrm{E},\mathrm{c}}-A_{\mathrm{P},\mathrm{c}}
  =
  (75.6\pm1.4)\,\mathrm{s}^{-1}.
\]

At this path-marked launch rate, Fig.~\ref{fig:results} shows no apparent
launch-induced change in D5's signal-singles fringe. The launch condition
retains the same small residual singles fringe common to all conditions, with
amplitude \(A_{\mathrm{L},s}=(22\pm3)\,\mathrm{s}^{-1}\), well below
\(R_{\mathrm{pm}}\). We attribute this common residual fringe (visibility
\(\approx 0.006\)) to imperfect polarization compensation in the
\(122\,\mathrm{m}\) fiber delay and other apparatus limitations.
\begin{figure}[t]
  \centering
  \includegraphics{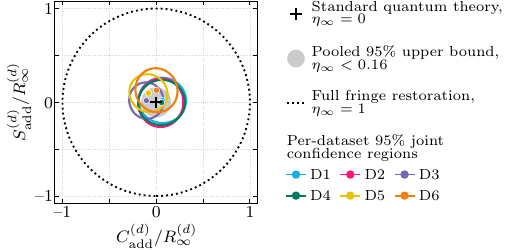}
  \caption{
    Cosine and sine quadratures of the launch-specific signal-singles interference
    fringe for datasets D1--D6, each normalized by its dataset-specific
    rate of path markers predicted to survive indefinitely.
    All exclude the full-restoration unit circle
    (\(\eta_\infty=1\)).
  }
  \label{fig:normalized-null-plane}
\end{figure}

The paired contrast fit to the 99\ matched launch--erase phase settings in
dataset D5 found no significant launch-specific fringe. Launch-only quadratures
\( (\hat C_{\mathrm{add}},\hat S_{\mathrm{add}}) = (-4.6\pm4.6,\,
5.4\pm4.6)\,\mathrm{s}^{-1} \) are consistent with zero, with amplitude \(\hat
A_{\mathrm{add}}=7.1\,\mathrm{s}^{-1}\) (Gaussian null \(p=0.33\); residual
sign-flip \(p=0.26\)). The 95\% amplitude bound was
\[
  A_{\mathrm{add},95}
  =18.8\,\mathrm{s}^{-1}.
\]
For dataset D5, the infinite-future survivor ensemble has a modeled
  path-marked rate of
\[
  R_\infty
  \equiv
  T_\infty R_{\mathrm{pm}}
  \approx
  56\,\mathrm{s}^{-1}.
\]
The broader finite-path ensemble requires survival only through Galactic
  transit, giving \( R_{\mathrm{fin}} \equiv T_{\mathrm{fin}}R_{\mathrm{pm}}
  \approx 56\,\mathrm{s}^{-1} \). The corresponding 95\% confidence bounds on
  restoration were
\[
  \eta_{\infty,95}\equiv\frac{A_{\mathrm{add},95}}{R_\infty}<0.34,
  \qquad
  \eta_{\mathrm{fin},95}\equiv\frac{A_{\mathrm{add},95}}{R_{\mathrm{fin}}}<0.34.
\]
On these normalizations, full restoration is excluded with nominal
  significance \(>5\sigma\) for both of dataset D5's ensembles and every phase.
  Further details for D5 are provided in
  Appendix~\ref{app:illustrative-dataset-case-study}.
  Table~\ref{tab:dataset-analysis} gives the rates of path-marked surviving
  idlers and exclusion bounds for each of the six datasets.

For the pooled analysis, we applied the same construction to all six datasets
and combined their launch-only quadrature estimates. The simulation-calibrated
pooled test was consistent with no added fringe (\(p=0.79\)). The fit of a
common restoration fraction gave \(\widehat\eta_\infty=0.09\) and
\(\eta_{\infty,95}<0.16\). This pooled bound and the normalized per-dataset
quadratures are shown in Fig.~\ref{fig:normalized-null-plane}. Full restoration
is excluded with nominal significance \(>5\sigma\). No between-dataset
heterogeneity was detected (\(p=0.95\)).

The finite-path normalization gives \(\eta_{\mathrm{fin},95}<0.16\) and
nominally excludes full restoration at \(>5\sigma\), essentially matching the
infinite-future result.

The exclusion holds for each of the six datasets individually, as shown in
Table~\ref{tab:dataset-analysis} and Fig.~\ref{fig:normalized-null-plane}.
Appendix~\ref{transmission-robustness} shows that the exclusion is robust to
large changes in the modeled transmission. The archived data and code reproduce
the complete analysis, from the experimental and environmental inputs to the
dataset-level and pooled results reported here~\cite{repo2026}.

\section{Discussion}
\label{sec:discussion}

The ensemble of idlers modeled to survive forever realizes the most stringent
operational form of path information available only ``in principle''; they
carry the sole path marker, remain perpetually unmeasured, and propagate
indefinitely. Together with the launch geometry, this precludes future causal
contact between these markers and their associated interferometric records. Our
null result is consistent with unconditioned signal interference depending
solely on joint-state path distinguishability, as quantum theory predicts.

Kent noted that the collapse-locality loophole may apply to
gravitational-collapse, spontaneous-localization, and consciousness-induced
collapse theories~\cite{Kent_2020}. Closing it requires causal separation only
until the relevant collapse or localization event, not forever. For an idler
that survives Galactic transit, a subsequent interaction could not influence
the laboratory for at least \(10^4\) years, far longer than any localization
interval. Normalization to this modeled ensemble gives essentially the same
bound as for the infinite-future ensemble, while relying only on the
well-constrained, finite-path portion of the transmission model.

\section{Alternatives}
\label{alternatives}

Simply declining to register the idler in a terrestrial apparatus does not
achieve causal separation. A beam dump, closed shutter, absorbing filter, or
unpowered detector does not remove the path marker; it transfers the encoded
path information to absorber degrees of freedom, thereby forming a causally
accessible environmental record.

Nor is finite storage equivalent to causal inaccessibility. A fiber loop,
optical cavity, or quantum memory can defer detection, but it keeps the marker
within the laboratory's causal domain. In practice, imperfect isolation during
storage can also allow some coupling between the marker and its environment.

\section{Prior work}
\label{sec:prior-work}

Photons lost from earlier experiments could in principle have entered
trajectories producing long or permanent causal separation. Ordinary
terrestrial losses, however, typically correspond to absorption by nearby
materials. Satellite uplinks are the most plausible accidental-escape channel:
photons that clear the atmosphere but miss the receiver aperture can continue
outward, including photons from ground-to-satellite entanglement-distribution
and teleportation
experiments~\cite{Han2020UplinkEntanglement,Ren2017Teleportation}. The reported
observables in those experiments were conditioned on ground--satellite
coincidence events, so photons that escaped without reaching the satellite
receiver were absent from the analyzed ensembles. More importantly, the
escaping photon was not deliberately used as the sole path marker for a partner
photon measured in a phase-scanned interferometer. Those data therefore did not
contain the unconditioned-singles observable needed to test complementarity or
the partial trace for an inaccessible marker in the setting of Kent's
loophole~\cite{Kent2005CausalQuantumTheory,Kent_2020}.

Salart \emph{et al.} addressed the essential collapse-locality loophole using
an \(18\,\mathrm{km}\) Bell-test baseline corresponding to
\(60\,\mu\mathrm{s}\) of light-travel time to separate putative
record-localization events under a Diósi--Penrose gravitational-collapse
criterion~\cite{Salart_2008}. Their Franson geometry suppressed single-photon
interference and provided no independent idler carrying which-path information
about the signal, so the unconditioned-singles partial-trace observable
considered here was absent.

Ag\"uero \emph{et al.} recently reported a variable-baseline Bell test with
stations separated by up to \(24\,\mathrm{m}\), corresponding to
\(80\,\mathrm{ns}\) of light-travel time~\cite{Aguero_2026}. Treating collapse
as an additional delay in the recorded detection time, they bounded its
duration by referencing photon time-stamps to classical trigger signals, then
checked that increasing the station separation produced neither an extra delay
nor a loss of the Bell violation. Their Bell-test observable was based on
two-photon correlations and did not involve a which-path marker or
interferometer; it did not test the partial-trace prediction for unconditioned
signal statistics.

Vedovato~\emph{et al.}~\cite{Vedovato2017SpaceDelayedChoice} extended Wheeler's
delayed-choice timing to satellite retroreflectors, testing the causal ordering
of interferometer choice and detection for a single-photon interference
observable rather than the fate of a partner photon carrying which-path
information.

Ma~\emph{et al.}~\cite{Ma_2013} implemented a delayed-choice quantum-eraser
test in which the erase-or-preserve choice and idler measurement were separated
from the signal detection by \(144\,\mathrm{km}\). In that protocol, the idler
was intentionally registered within approximately~\(0.5\,\mathrm{ms}\).
Rauch~\emph{et al.}~\cite{Rauch2018CosmicBell} used high-redshift quasars to
push the causal origin of Bell-test settings into the distant past. These prior
experiments causally separated settings, detections, or record-localization
events. The analyzed photons were measured, and their outcomes and settings
formed ordinary records within causal reach that were brought together for
comparison.

Yurtsever and Hockney proposed a closely related test in which one photon is
sent beyond a black-hole horizon~\cite{YurtseverHockney2005}. Here, surviving
idlers are modeled to pass beyond the laboratory's future cosmological event
horizon. In both, the possible signature is restored unconditioned singles
interference, violating the partial-trace identity that underlies no-signaling.

Partridge tested whether future absorption has present consequences, finding no
change in the power drawn by a microwave source radiating alternately into a
local absorber and into the open sky~\cite{Partridge1973AbsorberTheory}. Davies
later proposed an entangled-photon version using the retained partner's count
rate~\cite{Davies2014QuantumWeakMeasurements}. Neither involves a path marker
or interferometer.

Berera \emph{et al.} analyzed cosmological-distance quantum coherence and found
that soft X-ray photon mean free paths can be many orders of magnitude larger
than galactic scales, ``or even the observable
Universe''~\cite{Berera:2021xmn}. Modeling of a closed Friedmann universe
similarly suggests that initially optical radiation propagates essentially
without absorption until maximum expansion~\cite{DaviesTwamley1993,Craig1996}.

\section{Conclusion}
\label{sec:conclusion}

For the infinite-future survivor ensemble, the idler is modeled to remain
unmeasured and permanently causally inaccessible. To our knowledge, this is the
first experiment to monitor an observable of one member of an entangled pair in
this regime. The data showed no statistically significant launch-induced change
in unconditioned signal interference, consistent with quantum theory.

\begin{acknowledgments}
  We thank Jaime Calder{\'o}n-Figueroa for helpful comments that
  clarified our discussion of coherent propagation effects and the
  distinction between idler-state fidelity and path-marker
  distinguishability.
  We thank Arjun Berera for helpful correspondence regarding our discussion of
  Ref.~\cite{Berera:2021xmn}.
  We thank Pawan Gupta, principal investigator of the \ac{aeronet} site at the
  \ac{ucsb}, along with site manager Stuart Halewood and the site staff, for
  establishing and maintaining the site.
  We gratefully acknowledge support from the \ac{nsf} Quantum Foundry through
  the Q-AMASE-i program Award No. DMR-1906325 and the \ac{ucsb} \ac{nrt}
  Program Award No. 2152201.
\end{acknowledgments}

\runinheading{Data availability}
The data and analysis code that support the findings of this
article are openly available~\cite{repo2026}.

\appendix
\setcounter{topnumber}{2}
\setcounter{dbltopnumber}{2}

\section{Experimental details}
\label{sec:experimental-setup}

\subsection*{Source and detection electronics}
The continuous-wave \(405\,\mathrm{nm}\) pump power at the \ac{spdc} crystals
was \(28\,\mathrm{mW}\).

Photon detections were recorded with fiber-coupled silicon \ac{spad} modules.
The modules produced \(10\,\mathrm{ns}\), \(2.2\,\mathrm{V}\) \ac{ttl} pulses
with a specified dead time of \(24\,\mathrm{ns}\). Across the six datasets, the
largest observed raw singles rate was the preserve-condition idler rate,
approximately \(4.0\times10^4\,\mathrm{s}^{-1}\). With the specified
\(24\,\mathrm{ns}\) dead time, \(R\tau\approx9.7\times10^{-4}\) (less than
\(0.1\%\)), so dead-time corrections were neglected.

Detector clicks were recorded and coincidences were formed from detector
\ac{ttl} pulses using a full effective coincidence-window width of
\(w=25\,\mathrm{ns}\). During each \(10\,\mathrm{s}\) acquisition at a fixed
\ac{mzi} phase setting, the \ac{cc} reported aggregate singles counts
\(N_{S_1},N_{S_2},N_I\) from the two signal detectors and the idler detector,
together with the corresponding signal--idler coincidence counts
\(N_{C_1},N_{C_2}\). In the launch condition, only \(N_{S_1}\) and \(N_{S_2}\)
were analyzed because the idler was not detected.

\subsection*{Fiber polarization compensation}
The \(122\,\mathrm{m}\) signal delay fiber and the \(61\,\mathrm{m}\) idler
roof fiber apply slowly drifting unitary polarization transformations. These
must be compensated so that the \ac{mzi} input \ac{pbs} splits the signal path
alternatives in the source \(\mathrm H/\mathrm V\) basis and so that the
coincidence fringe visibility in the erase condition is maximized at the fixed
roof eraser polarizer. Each fiber was preceded by an in-line three-paddle
polarization controller, equivalent to three adjustable wave plates. The six
paddle angles were tuned automatically by online Bayesian optimization using a
Gaussian process model~\cite{Wigley2016MLOOP} of an experimental figure of
merit.

Each evaluation of the figure of merit combined polarimetric and
interferometric measurements. For the polarimetric part, signals were prepared
in horizontal and diagonal polarizations using the calibration \ac{hwp}
upstream of the delay fiber and were analyzed after the fiber by rotating the
\ac{hwp} immediately before the \ac{mzi} input \ac{pbs}. Together, this
\ac{hwp} and the \ac{pbs} act as a rotatable linear analyzer. During these
scans, the beam block interrupted one \ac{mzi} arm to prevent interference, and
the coincidence rates were fit to Malus's law. Each fit gave the transmitted
polarization axis \(\theta_0\) and the Malus-fringe contrast \(k\), which
together define the equatorial Stokes components
\((s_1,s_2)=k(\cos2\theta_0,\sin2\theta_0)\). The figure of merit penalized the
Euclidean distances between these vectors and the targets \((1,0)\) and
\((0,1)\) on the Poincar\'e equator. Scan durations were extended adaptively
until the fitted uncertainties reached \(1^\circ\) in \(\theta_0\) and
\(0.025\) in \(k\).

The idler controller requires no absolute polarization reference. The eraser
polarizer on the roof is fixed, so it suffices for the roof fiber to transform
the two idler marker states so that the fixed polarizer projects them with
equal weight, which maximizes the coincidence fringe visibility in the erase
condition. The interferometric term of the figure of merit provides exactly
this criterion. As part of each figure-of-merit evaluation, an erase-condition
\ac{mzi} phase scan measured the coincidence fringe amplitude
\(A_{\mathrm{c}}\) and the residual unconditioned signal-singles fringe
amplitude \(A_{\mathrm{s}}\) and contributed the contrast
\((A_{\mathrm{c}}-A_{\mathrm{s}})/(A_{\mathrm{c}}+A_{\mathrm{s}})\) with triple
weight relative to each polarimetric distance. This term serves two purposes.
It is the sole tuning criterion for the idler controller, and it simultaneously
penalizes any residual signal-singles fringe from imperfect signal-side
compensation.

\subsection*{Paired \ac{mzi} phase acquisitions}

The fixed voltage grid was constructed from the nominal fringe period
\(P_{\mathrm{nom}}=2.05\,\mathrm{V}\). This nominal value defined only the
acquisition grid and was not imposed as the period in the fringe analysis. A
half-open window of width \(1.5P_{\mathrm{nom}}\), sampled at intervals of
\(P_{\mathrm{nom}}/7\), gave 11 settings. Within each of nine\ scans, settings
were visited in recursive bisection order so that early acquisitions sampled
the full scan interval. At each setting, the paired conditions were measured
consecutively for \(10\,\mathrm{s}\) each. The setting visit order and
condition order were reversed on alternating scans, giving \(90\,\mathrm{s}\)
per condition and setting over each run.
\subsection*{Dark-run backgrounds and accidental coincidences}
For each condition, background measurements were taken with the pump laser
blocked; they used the same channels and full effective coincidence-window
width \(w=25\,\mathrm{ns}\) as the corresponding acquisitions. For each
acquisition \(k\), with duration \(\Delta t=10\,\mathrm{s}\), we defined
\[
  \begin{aligned}
    d_X
     & = \frac{N_X^{\mathrm{dark}}}{\Delta t_{\mathrm{dark}}},    \\
    N_{\mathrm{acc},j,k}
     & = \frac{w}{\Delta t}N_{I,k}N_{S_j,k},                      \\
    b_{C_j}
     & = d_{C_j}-w\,d_I d_{S_j},                                  \\
    R_{C_j,k}^{\mathrm{corr}}
     & = \frac{N_{C_j,k}-N_{\mathrm{acc},j,k}}{\Delta t}-b_{C_j}.
  \end{aligned}
\]
Here \(N_X^{\mathrm{dark}}\) is the count accumulated on channel \(X\) during
  the condition-matched dark acquisition, and \(d_X\) is the corresponding
  in-apparatus background rate, not an intrinsic detector dark-count rate. It
  may include intrinsic \ac{spad} dark counts, optical or ambient background,
  and electronic background. The term \(w d_I d_{S_j}\) removes dark--dark
  coincidences already included in the pointwise accidental estimate and thus
  prevents double subtraction. Uncertainties on \(d_X\) are Poisson standard
  deviations, \(\sqrt{N_X^{\mathrm{dark}}}/\Delta t_{\mathrm{dark}}\), and
  those on \(b_{C_j}\) are propagated from the measured dark counts.

The nonnegative, dark-subtracted signal-singles rates used for fringe plotting
and the paired launch--erase contrast fit were
\[
  R_{S_j,k}^{\mathrm{corr}}
  = \max\left(\frac{N_{S_j,k}}{\Delta t}-d_{S_j},0\right).
\]
Raw rates were formed from total counts and total acquisition time.
  Accidentals and corrected rates were evaluated pointwise and then averaged
  over acquisitions. The coincidence totals are sums over the two signal
  outputs, \(R_C^{\mathrm{raw}}=\sum_j R_{C_j}^{\mathrm{raw}}\) and
  \(R_C^{\mathrm{corr}}=\sum_j R_{C_j}^{\mathrm{corr}}\), and
  \(f_{\mathrm{acc}}=R_{\mathrm{acc}}/R_C^{\mathrm{raw}}\).

\section{Transmission calculations}
\label{app:transmission-details}

We count a launched idler as surviving over a propagation interval when it
remains an unmeasured carrier of the polarization path marker. The idler must
avoid absorption and scattering over that interval. Any other interactions must
preserve the overlap of the marker states and deposit no path information in an
environmental record.

Survival does not require the idler to maintain fidelity to the optical
wavepacket emitted at the source. The signal-singles visibility depends on the
overlap of the two idler marker states, $|\langle m_{\mathrm U}|m_{\mathrm
L}\rangle|$. Coherent propagation that changes the detailed photon state is
not, by itself, a failed survival event. Redshift, diffraction, lensing, phase
shifts, parallel transport, and lossless polarization evolution can change the
idler wavepacket, and hence its fidelity to the launched state, without
changing the marker overlap. If the surviving idler undergoes a unitary $U$ on
its spatial, spectral, and polarization degrees of freedom, then $$ \langle
m_{\mathrm U}'|m_{\mathrm L}'\rangle = \langle m_{\mathrm U}|U^\dagger
U|m_{\mathrm L}\rangle = \langle m_{\mathrm U}|m_{\mathrm L}\rangle . $$ The
surviving-marker analysis is altered only if the idler undergoes a nonunitary,
marker-dependent process. Examples include polarization-dependent loss,
filtering, depolarization, or an interaction that leaves an environmental
record correlated with the marker.

We apply this survival criterion to two launch-condition reference ensembles.
For the finite-path ensemble, the propagation interval extends through the
Milky Way. For the infinite-future ensemble, it extends over the idler's entire
future trajectory.

Both ensembles therefore require the idlers to pass coherently through the
launch optics, the atmosphere, and the Milky Way without creating an
environmental path record. The infinite-future ensemble additionally requires
coherent propagation through the expanding \ac{igm}, modeled to \(a\to\infty\).
Across the six launch runs, \(T_{\mathrm{fin}}\) ranges from \(0.75\) to
\(0.83\), while \(T_\infty\) ranges from \(0.73\) to \(0.82\).

Table~\ref{tab:dataset-transmission} gives the dataset-specific atmospheric and
Galactic factors and the resulting transmissions for both ensembles. For each
dataset, both transmissions use the minimum sampled Galactic transmission over
the launch acquisition. The launch-optics factor is common to both ensembles
and all datasets. The intergalactic factor is common across datasets but enters
only \(T_\infty\). Detailed calculations for each component follow.

\begin{table*}
  \caption{
    Transmission factors for all six datasets.
    The first column gives dataset identifiers D1--D6 in chronological order.
    Superscripts \(\mathrm{G}\) and \(\mathrm{A}\) identify
    \acl*{goes18} and lunar \acl*{aeronet} inputs, respectively.
    \(T_{\mathrm{MW}}(t_0)\) is evaluated at the start of the launch run,
    \(\langle T_{\mathrm{MW}}\rangle\) is the sampled
    mean over the 49-minute wall-clock launch-run interval, and
    \(\min(T_{\mathrm{MW}})\) is the minimum sampled value
    over that interval.
    For the analysis reported here, both transmissions use this minimum,
    \(T_{\mathrm{fin}}=T_{\mathrm{opt}}T_{\mathrm{atm}}
    \min(T_{\mathrm{MW}})\) and
    \(T_\infty=T_{\mathrm{fin}}T_{\mathrm{IGM},\infty}\).
    The common factors are
    \(T_{\mathrm{opt}}=0.9853\) and
    \(T_{\mathrm{IGM},\infty}=0.9839\).
    Extra digits are carried for arithmetic reproducibility, not as
    significant figures.
  }
  \label{tab:dataset-transmission}
  \centering
  \begin{tabular*}{\linewidth}{@{}l@{\extracolsep{\fill}}cccccc@{}}
    \toprule
    & \multicolumn{6}{c}{Transmission} \\
    \cmidrule(l){2-7}
    Dataset
    & $T_{\mathrm{atm}}$
    & $T_{\mathrm{MW}}(t_0)$
    & $\langle T_{\mathrm{MW}}\rangle$
    & $\min(T_{\mathrm{MW}})$
    & $T_{\mathrm{fin}}$
    & $T_\infty$ \\
    \midrule
    D1 & $0.9126^{\mathrm{G}}$ & $0.9273$ & $0.9013$ & $0.8449$ & $0.7597$ & $0.7474$ \\
    D2 & $0.8821^{\mathrm{G}}$ & $0.9028$ & $0.8981$ & $0.8606$ & $0.7480$ & $0.7359$ \\
    D3 & $0.8982^{\mathrm{A}}$ & $0.9384$ & $0.9594$ & $0.9384$ & $0.8305$ & $0.8171$ \\
    D4 & $0.8623^{\mathrm{G}}$ & $0.9781$ & $0.9593$ & $0.9436$ & $0.8017$ & $0.7888$ \\
    D5 & $0.8873^{\mathrm{G}}$ & $0.8543$ & $0.8836$ & $0.8543$ & $0.7468$ & $0.7348$ \\
    D6 & $0.9151^{\mathrm{G}}$ & $0.9358$ & $0.9011$ & $0.8654$ & $0.7803$ & $0.7677$ \\
    \bottomrule
  \end{tabular*}
\end{table*}

We model transmission across a wavelength range around the nominal 810\,nm pair
wavelength. The signal photons in the lab had to pass an $800\pm40$\,nm
interference filter to be detected by the \acp{spad} at the exit of the
\ac{mzi}. The launched idlers were filtered by nothing narrower than a 610\,nm
long-pass filter, but only a narrow-band subset is relevant here. We can
estimate the wavelengths of the idlers paired with signals that passed the
\ac{spad} interference filters. For each detected signal wavelength
$\lambda_s$, the paired idler wavelength is set by \ac{spdc} energy
conservation
\[
  \frac{1}{\lambda_p}=\frac{1}{\lambda_s}+\frac{1}{\lambda_i},
  \qquad
  \lambda_p=405\,\mathrm{nm}.
\]

The relevant launched idlers are thereby frequency-anticorrelated with the
detected signals and occupy a band spanning approximately 782--867\,nm.

\subsection*{Launch optics}

Before entering the open atmosphere, each launched idler reflected from the
upward-turning launch mirror and passed through the silica launch window. We
treat failure of specular reflection from the mirror, and reflection,
absorption, or scattering at the window, as failed survival events. Such
photons are no longer in the intended outgoing launch mode and can leave path
records in the environment.

The launch mode had a \(1/e^2\) beam diameter of roughly \(2.4\,\mathrm{mm}\)
at the mirror. The corresponding diffraction-limited full-angle divergence was
about \(0.4\,\mathrm{mrad}\). Beam expansion after launch is treated as unitary
propagation, not as loss or path-record formation.

The launch mirror was a Thorlabs BB1-E03 broadband dielectric mirror used at an
angle of incidence of \(45^\circ\). The manufacturer specifies
\(R_{\mathrm{avg}}>99.2\%\) for both \(s\)- and \(p\)-polarized light over
\(780\)--\(870\,\mathrm{nm}\), covering the 782--867\,nm idler band relevant
for detected signals. We therefore use the conservative specified lower bound
\[
  T_{\mathrm{mir}}=0.992.
\]

The roof window was a Thorlabs WG41050-B \ac{uvfs} window with the B broadband
antireflection coating on both surfaces. The specified average reflectance is
\(R_{\mathrm{surf}}<0.0034\) per surface at near-normal incidence over the same
wavelength band. Bulk absorption in \(5\,\mathrm{mm}\) of \ac{uvfs} at
\(810\,\mathrm{nm}\) is negligible, so the no-reflection window transmission is
\[
  T_{\mathrm{win}}
  =
  (1-R_{\mathrm{surf}})^2
  \ge
  (1-0.0034)^2
  =
  0.9932.
\]
The outside of the window was protected with a lens cap except during runs
  that included the launch condition. Before each launch run, the window was
  visually inspected and cleaned with a residue-free air duster.

The launch-optics transmission used in the survival budget is therefore
\[
  T_{\mathrm{opt}}
  \equiv
  T_{\mathrm{mir}}T_{\mathrm{win}}
  \ge
  0.992(1-0.0034)^2
  =
  0.9853.
\]
In the numerical budget, we conservatively set \(T_{\mathrm{opt}}\) to the
  specified lower bound \(0.9853\), treating the corresponding launch-optics
  loss as record-forming.

\subsection*{Atmospheric transmission}

We estimate the atmospheric transmission using libRadtran, a library for
radiative transfer~\cite{Emde2016libRadtran,Mayer2005libRadtran}, including the
representative wavelengths band parameterization~\cite{Gasteiger2014REPTRAN}.
We use a clear-sky, zenith, direct-beam transmittance calculation.

For each dataset, we used atmospheric observations near the rooftop launch site
at Henley Hall on the \ac{ucsb} campus. Henley Hall is situated \(\sim
1.1\)\,km from the Santa Barbara Municipal Airport (KSBA) automated surface
observing station, which issues frequent \acp{metar} of current surface
conditions. Atmospheric-column inputs were taken from either the \ac{goes18}
weather satellite operated by the \ac{noaa}~\cite{GOES18AOD,GOES18TPW} or lunar
\ac{aeronet} observations~\cite{Holben1998,Giles2019,Schafer2026LunarAERONET},
depending on the dataset. For satellite-based datasets, we combined
conservatively selected inputs from screened \ac{goes18} observations within
\(5\,\mathrm{km}\) of the launch site. This bounds spatial variability
conservatively while drawing only on high-quality retrievals, since the
\ac{goes18} aerosol algorithms are validated over land and open water and flag
mixed coastal pixels as lower quality.

All six datasets were acquired under clear-sky conditions, as confirmed by
direct visual observation and contemporaneous meteorological data. For each
launch acquisition, the nearest archived KSBA\ \ac{metar} reported
CLR~\cite{AviationWeatherMETAR_20260305_08}, indicating that no clouds were
detected below approximately \(3.7\)\,km (\(12\)\,kft). For the five daytime
datasets, the \ac{goes18} \ac{cod} indicated no clouds~\cite{GOES18COD}.

Atmospheric transmission varies sharply for wavelengths near 810\,nm due to the
presence of nearby oxygen and water absorption bands. The launched-idler band
spans approximately 782--867\,nm and overlaps the water absorption bands at
$\sim$ 820--850\,nm, but not the $\sim$~758--775\,nm oxygen bands.

For the libRadtran calculation, we use a top-hat (rectangular) filter and
report the resulting averaged transmission
\[
  T_{\mathrm{atm}} =
  \frac{\int T(\lambda)\,F(\lambda)\,d\lambda}{\int F(\lambda)\,d\lambda},
\]
where $F(\lambda)=1$ for $782\le \lambda \le 867$\,nm and zero otherwise,
  representing the relevant idler bandwidth.
\subsection*{Galactic transmission}
The amount of the Milky Way's dusty bulk traversed by vertically launched
photons depends on the zenith launch direction's position relative to the
Galactic plane at the time of launch. We estimate Galactic attenuation via
\[
  T_{\mathrm{MW}} = 10^{-0.4\,A_\lambda},
\]
\[
  \tau_{\mathrm{MW}} \equiv -\ln T_{\mathrm{MW}} = 0.4\,\ln 10\,A_\lambda.
\]
We utilize the geometric zenith over \ac{ucsb} to query the \ac{irsa}
  Galactic DUST Reddening and Extinction service~\cite{irsa_dust}. \Ac{irsa}
  provides $A_\lambda$ for $\lambda=811.1$\,nm, which we use to model the
  nominal idler wavelength of 810\,nm. Although \ac{irsa} tabulates extinction
  for extragalactic light arriving at Earth, the corresponding outbound optical
  depth is the same, so we use these data to estimate the idler loss.

We adopted the minimum sampled Galactic transmission during each launch window.
The Milky Way moved across the sky during each launch session, gradually
altering transmission. For comparison, we also provide the sampled mean
transmission over each launch window.

We define the finite-path ensemble as idlers that remain unmeasured carriers of
the path marker over the first \(10^4\) light-years of propagation. At this
distance from Earth, the launch trajectories lie approximately
\(1\)–\(2\,\mathrm{kpc}\) from the Galactic plane, well beyond the main
Galactic dust layer~\cite{Schlafly_2011}. Its modeled transmission is
\(T_{\mathrm{fin}}\equiv T_{\mathrm{opt}}T_{\mathrm{atm}}T_{\mathrm{MW}}\). An
interaction after this interval could influence the laboratory only after a
round-trip light-travel time of at least \(2\times10^4\) years. We quote the
weaker lower bound \(10^4\) years.

\subsection*{Intergalactic transmission}

After the finite atmospheric and Galactic paths, the idler enters the expanding
\ac{igm}. We normalize an effective present-day dust opacity from a
low-redshift inference, add conservative galaxy-interception and free-electron
scattering terms, and evolve all three along the photon's future null
trajectory. In the integral, proper densities dilute with expansion and the
photon wavelength redshifts as \(\lambda(a)=\lambda_0 a\). This redshift
changes the spectral wavepacket used in the opacity calculation, but because it
is common to the two polarization-marker alternatives, it does not change the
marker-state overlap.

We integrate the resulting optical depth to \(a\to\infty\). Dust extinction and
galaxy interception make comparable contributions, while Thomson scattering by
free electrons gives a smaller correction,
\[
  \tau_{\mathrm{IGM},\infty}
  =
  \tau_{\mathrm{dust},\infty}
  +\tau_{\mathrm{gal},\infty}
  +\tau_{e,\infty}.
\]
For a flat \ac{lambdacdm} model with $a=1$ at launch, we use
  \(H_0=67.4\,\mathrm{km\,s^{-1}\,Mpc^{-1}}\), \(\Omega_m=0.315\), and
  \(\Omega_\Lambda=0.685\) from Planck 2018~\cite{planck:2018} and define
\[
  E(a)=\frac{H(a)}{H_0}=\sqrt{\Omega_m a^{-3}+\Omega_\Lambda}.
\]
Along the idler trajectory,
\[
  dt=\frac{da}{aH_0E(a)},
  \qquad
  \lambda(a)=\lambda_0 a.
\]
For the dust term, Ménard \emph{et al.} inferred \(A_V(z=0.5)\approx0.03\)
  mag over approximately \(1.9\)\,Gpc of comoving
  distance~\cite{Menard2010IGMDust}. This is a model-dependent observer-frame
  integral of opacity correlated with galactic halos and large-scale structure,
  not a direct measurement of a present-day local physical extinction
  coefficient. We divide by the comoving distance only to construct an
  effective mean cosmic coefficient. Converting magnitudes to optical depth
  gives the visual-band normalization
\[
  k_{V,0}=0.4\ln(10)\,\frac{0.03\,\mathrm{mag}}{1.9\,\mathrm{Gpc}}
  \approx1.45\times10^{-2}\,\mathrm{Gpc^{-1}}.
\]

Xie \emph{et al.}~\cite{xie:2015} modeled the redshift dependence of quasar
continua at \(z<1.5\) and obtained a comparable visual-band optical-depth
coefficient of \(1.3\times10^{-2}\,\mathrm{Gpc}^{-1}\) when scaled to our
adopted \(H_0\). This value is comparable in scale to the effective coefficient
adopted above from Ménard et al.

We convert this normalization to the launch wavelength
\(\lambda_0=810\,\mathrm{nm}\) using the Cardelli, Clayton, and Mathis
extinction curve with \(R_V=3.1\)~\cite{cardelli:1989},
\[
  k_{810,0}=k_{V,0}\frac{A_{810\,\mathrm{nm}}}{A_V}
  \approx8.48\times10^{-3}\,\mathrm{Gpc^{-1}}.
\]
In the integral below, \(k_{810,0}\) is interpreted as the present-epoch
  physical extinction coefficient per proper Gpc.

Let $g(a)$ denote the relative dust opacity at the redshifted wavelength
$\lambda_0a$, normalized so that $g(1)=1$. We compute $g(a)$ using the same
extinction curve, extended with its near-\ac{ir} power-law behavior at longer
wavelengths. The far-\ac{ir} tail has little effect because the density
dilution has already suppressed the integrand by the time the idler reaches
those wavelengths.

With \(c\) the speed of light, the dust optical depth is
\[
  \tau_{\mathrm{dust},\infty}
  =
  k_{810,0}\frac{c}{H_0}
  \int_1^\infty \frac{g(a)}{a^4E(a)}\,da.
\]
Numerically,
\[
  \begin{aligned}
    \int_1^\infty \frac{g(a)}{a^4E(a)}\,da
     & = 0.227,             \\
    \tau_{\mathrm{dust},\infty}
     & = 8.57\times10^{-3}.
  \end{aligned}
\]

Craig's conservative galaxy-interception model treats each galaxy as an opaque
disk~\cite{Craig1996}. We adopt
\[
  n_{\mathrm{gal}}(a)=n_{\mathrm{gal},0}a^{-3},
  \qquad
  \sigma_{\mathrm{gal}}=\pi r_{\mathrm{gal}}^2,
\]
with \(n_{\mathrm{gal},0}=0.02h^3\,\mathrm{Mpc^{-3}}\) and
  \(r_{\mathrm{gal}}=10^4h^{-1}\,\mathrm{pc}\), where
  \(h=H_0/(100\,\mathrm{km\,s^{-1}\,Mpc^{-1}})\). Every trajectory crossing a
  projected disk is counted as lost or as forming an environmental record, even
  though a real galaxy need not be opaque over its entire disk. Assuming a
  fixed physical radius and constant comoving abundance gives
\[
  \begin{aligned}
    \tau_{\mathrm{gal},\infty}
                            & =
    n_{\mathrm{gal},0}\pi r_{\mathrm{gal}}^2\frac{c}{H_0}
    \int_1^\infty\frac{da}{a^4E(a)}
    =6.87\times10^{-3},                \\
    T_{\mathrm{gal},\infty} & =0.9932.
  \end{aligned}
\]
This galaxy-interception estimate is independent of \(h\), whose factors
  cancel in \(n_{\mathrm{gal},0}\sigma_{\mathrm{gal}}H_0^{-1}\). Because the
  calculation treats every disk as fully opaque and holds the comoving galaxy
  population fixed into the future, we use it as a conservative upper estimate.

The proper free-electron density evolves as $n_e(a)=n_{e,0}a^{-3}$. We take
$n_{e,0}$ to be the present-day mean baryon density,
$n_b\approx2.5\times10^{-7}\,\mathrm{cm^{-3}}$, consistent with Planck
cosmological parameters~\cite{planck:2018}. For simplicity, the model uses the
conservative upper estimate $n_{e,0}\approx n_b$, corresponding to one free
electron per baryon. This is conservative because a fully ionized primordial
hydrogen-helium mixture gives about \(0.88\) free electrons per baryon, and
some baryons are not in the diffuse ionized \ac{igm}. For a Thomson cross
section of $\sigma_T = 6.652\times10^{-25}\,\mathrm{cm^2}$, the corresponding
optical depth is
\[
  \tau_{e,\infty}
  =
  n_{e,0}\sigma_T\frac{c}{H_0}
  \int_1^\infty \frac{da}{a^4E(a)}.
\]
Numerically,
\[
  \int_1^\infty \frac{da}{a^4E(a)}=0.365.
\]
This corresponds to an effective present-density path length of
  \(1.62\,\mathrm{Gpc}\) and gives
\[
  \tau_{e,\infty}=8.33\times10^{-4},
  \qquad
  T_{e,\infty}=0.9992.
\]

Combining all three terms gives
\[
  \begin{aligned}
    \tau_{\mathrm{IGM},\infty}
     & = 1.63\times10^{-2},                \\
    T_{\mathrm{IGM},\infty}
     & = \exp[-\tau_{\mathrm{IGM},\infty}]
    = 0.9839.
  \end{aligned}
\]

To test sensitivity to wavelength-dependent extinction, we retain the same
launch opacity \(k_{810,0}\) and set \(g(a)=1\) for all future wavelengths,
while retaining cosmological density dilution. This gives
\[
  \tau_{\mathrm{dust},\infty}^{g=1}
  = 1.38\times10^{-2}.
\]
Including galaxy interception and electron scattering gives
  \(\tau_{\mathrm{IGM},\infty}^{g=1} =2.15\times10^{-2}\) and
  \(T_{\mathrm{IGM},\infty}^{g=1}=0.9788\). Thus the conclusion that
  intergalactic attenuation is a percent-level correction is insensitive to the
  modeled decline in extinction as the idler redshifts.

The integral converges rapidly. At late times $E(a)\to\sqrt{\Omega_\Lambda}$,
so the galaxy and electron integrands fall as $a^{-4}$; the dust integrand
falls at least this fast. Accelerated expansion also makes the future conformal
distance finite.

Table~\ref{tab:igm-infinity} summarizes the input quantities and derived
transmission factors for the \ac{igm} calculation.

\subsection*{Other attenuation and record-forming channels}

This budget accounts for the standard interactions expected to contribute
appreciably to the optical depth of future-propagating \(810\,\mathrm{nm}\)
idlers. Additional effects along the future trajectory are either kinematically
inaccessible, negligibly weak, or coherent rather than decohering.

Lyman-$\alpha$ absorption and ground-state bound-free photoionization are
kinematically excluded because the idlers enter the \ac{igm} at near-\ac{ir}
wavelengths and subsequently redshift only to longer wavelengths. Gas Rayleigh
scattering and free--free absorption are allowed in principle but have optical
depths far below even the Thomson term. Molecular line absorption is likewise
expected to be negligible for the diffuse \ac{igm} because the relevant
molecular abundances and covering factors are very small.

Other possible cosmological attenuation or record-forming channels are not
included in the budget because they are expected to be negligible here.
Photon--photon scattering on the cosmic microwave background or extragalactic
background light is expected to be far below the modeled \ac{igm} terms, and
high-energy particle-production channels are kinematically inaccessible.
Quantum photon--graviton scattering could in principle entangle the photon
polarization with gravitational degrees of freedom and thereby form a record.
Such interactions are expected to be far below the modeled \ac{igm} terms.

Coherent plasma, magnetic, birefringent, or gravitational propagation effects
may rotate the idler polarization, shift its phase or frequency, or otherwise
change the photon wavepacket, but such unitary propagation does not change the
marker-state overlap.

\begin{table}[t]
  \caption{
    Intergalactic transmission common to all six datasets, evaluated
    along the idler’s future null trajectory to \(a\to\infty\).
    The table lists the inputs and derived \acf*{igm} transmission factors.
    Nominal values are carried for arithmetic reproducibility, not significant figures.
  }
  \label{tab:igm-infinity}
  \centering
  \begin{tabular*}{\columnwidth}{@{}l@{\extracolsep{\fill}}cr@{}}
    \toprule
    Quantity & Source & Value \rule[-1ex]{0pt}{0pt}\\
    \midrule
    $H_0$ & Planck 2018 & $67.4\,\mathrm{km\,s^{-1}\,Mpc^{-1}}$ \rule{0pt}{2.5ex}\\
    $\Omega_m$ & Planck 2018 & $0.315$ \\
    $\Omega_\Lambda$ & Planck 2018 & $0.685$ \\
    $A_V(z=0.5)$ & \cite{Menard2010IGMDust} & $\approx0.03$ mag over $\sim1.9$\,Gpc \\
    $k_{V,0}$ & Derived & $1.45\times10^{-2}\,\mathrm{Gpc^{-1}}$ \\
    $A_{810\,\mathrm{nm}}/A_V$ & \cite{cardelli:1989} & $0.583$ \\
    $k_{810,0}$ & Derived & $8.48\times10^{-3}\,\mathrm{Gpc^{-1}}$ \\
    $n_{\mathrm{gal},0}$ & \cite{Craig1996} & $0.02h^3\,\mathrm{Mpc^{-3}}$ \\
    $r_{\mathrm{gal}}$ & \cite{Craig1996} & $10^4h^{-1}\,\mathrm{pc}$ \\
    $n_b$ & Planck 2018 & $2.5\times10^{-7}\,\mathrm{cm^{-3}}$ \\
    $n_{e,0}$ & Model & $\approx n_b$ \\
    $\sigma_T$ & Constant & $6.652\times10^{-25}\,\mathrm{cm^2}$ \\
    \midrule
    $\tau_{\mathrm{dust},\infty}$ & Derived & $8.57\times10^{-3}$ \\
    $T_{\mathrm{dust},\infty}$ & Derived & $0.9915$ \rule[-1ex]{0pt}{0pt}\\
    \midrule
    $\tau_{\mathrm{gal},\infty}$ & Derived & $6.87\times10^{-3}$ \\
    $T_{\mathrm{gal},\infty}$ & Derived & $0.9932$ \rule[-1ex]{0pt}{0pt}\\
    \midrule
    $\tau_{e,\infty}$ & Derived & $8.33\times10^{-4}$ \\
    $T_{e,\infty}$ & Derived & $0.9992$ \rule[-1ex]{0pt}{0pt}\\
    \midrule
    $\tau_{\mathrm{IGM},\infty}$ & Derived & $1.63\times10^{-2}$ \rule{0pt}{2.5ex}\\
    $T_{\mathrm{IGM},\infty}$ & Derived & $0.9839$ \\
    \bottomrule
  \end{tabular*}
\end{table}

\subsection*{Prior cosmological propagation studies}

Berera \emph{et al.} analyzed potential decoherence mechanisms for photons
propagating over cosmological distances~\cite{Berera:2021xmn}. Their soft
X-ray, high-redshift analysis does not apply to our future-directed
near-\ac{ir} idlers, but it is a useful precedent for cosmological-scale
coherence.

The small integrated future opacity also has precedent in studies of
cosmological future-light-cone
opacity~\cite{Davies1972,DaviesTwamley1993,Craig1996}. Davies gave a general
criterion for complete cosmological absorption, while Davies and Twamley, and
later Craig, evaluated propagation to maximum expansion in closed Friedmann
models. Craig obtained an upper-limit optical depth of order \(10^{-2}\) at
optical frequencies. As in the calculation above, cosmological dilution of
absorbers outweighs the increasing propagation distance.

\FloatBarrier
\section{Launched-idler normalization}

\begin{table*}
  \caption{
    Joint interference-fringe fit parameters for each experimental condition
    of the illustrative dataset D5.
    Uncertainties are marginal \(1\sigma\) uncertainties from the final fit
    covariance; for the derived amplitude \(A\), the full fit covariance is
    propagated. Extra digits are retained for arithmetic reproducibility, not
    as significant figures.
  }
  \label{tab:fits}
  \centering
  \begin{tabular}{lllrrrr}
    \toprule
    Run & Condition & Detection channel & $\bar R_1$ ($\mathrm{s}^{-1}$) & $\bar R_2$ ($\mathrm{s}^{-1}$) & $\mathcal{V}$                & $A$ ($\mathrm{s}^{-1}$)   \\
    \midrule
    1   & Launch    & Signal singles    & 2041.4 \ensuremath{\pm} 3.5    & 1596.8 \ensuremath{\pm} 3.2    & 0.006 \ensuremath{\pm} 0.001 & 22.1 \ensuremath{\pm} 3.1 \\
    1   & Erase     & Coincidences      & 78.4 \ensuremath{\pm} 0.5      & 64.4 \ensuremath{\pm} 0.5      & 0.567 \ensuremath{\pm} 0.006 & 81.0 \ensuremath{\pm} 1.0 \\
    1   & Erase     & Signal singles    & 2037.3 \ensuremath{\pm} 3.5    & 1597.9 \ensuremath{\pm} 3.2    & 0.007 \ensuremath{\pm} 0.001 & 23.7 \ensuremath{\pm} 3.2 \\
    2   & Preserve  & Coincidences      & 234.7 \ensuremath{\pm} 0.6     & 183.0 \ensuremath{\pm} 0.5     & 0.013 \ensuremath{\pm} 0.002 & 5.4 \ensuremath{\pm} 1.0  \\
    2   & Preserve  & Signal singles    & 2026.1 \ensuremath{\pm} 3.5    & 1589.8 \ensuremath{\pm} 3.2    & 0.005 \ensuremath{\pm} 0.001 & 18.5 \ensuremath{\pm} 3.1 \\
    2   & Erase     & Coincidences      & 77.8 \ensuremath{\pm} 0.6      & 63.5 \ensuremath{\pm} 0.5      & 0.574 \ensuremath{\pm} 0.006 & 81.1 \ensuremath{\pm} 1.0 \\
    2   & Erase     & Signal singles    & 2025.8 \ensuremath{\pm} 3.5    & 1588.6 \ensuremath{\pm} 3.2    & 0.005 \ensuremath{\pm} 0.001 & 17.3 \ensuremath{\pm} 3.1 \\
    \bottomrule
  \end{tabular}
\end{table*}
\label{sec:physical-ensemble}%
\(R_{\mathrm{pm}}\), and hence \(R_\infty\) and \(R_{\mathrm{fin}}\), is
derived from coincidence amplitudes conditioned on transmission through the
eraser polarizer and idler detection. It therefore undercounts the coherent
launched-idler rate. The launch, erase, and preserve configurations share the
idler path up to the roof flip mount. Their differences are downstream, with
launch-unique mirror, window, and sky losses already included in the
transmission factors.

Let \(\mathcal A_{\mathrm{E,c}}\) and \(\mathcal A_{\mathrm{P,c}}\) be the
complex two-output coincidence fringe amplitudes in the erase and preserve
conditions. The fits report their magnitudes, \(A_{\mathrm{E,c}}=|\mathcal
A_{\mathrm{E,c}}|\) and \(A_{\mathrm{P,c}}=|\mathcal A_{\mathrm{P,c}}|\). Let
\(\mathcal A_{\mathrm{comp,c}}\) denote the inferred complex fringe
contribution from the analyzer outcome complementary to the eraser's accepted
projection. Linearity of the conditioned signal coherence gives
\[
  \mathcal A_{\mathrm{P,c}}
  =
  \mathcal A_{\mathrm{E,c}}
  +
  \mathcal A_{\mathrm{comp,c}}.
\]
Hence
\begin{align*}
  |\mathcal A_{\mathrm{comp,c}}|
   & =|\mathcal A_{\mathrm{P,c}}-\mathcal A_{\mathrm{E,c}}| \\
   & \ge
  \bigl||\mathcal A_{\mathrm{P,c}}|
  -|\mathcal A_{\mathrm{E,c}}|\bigr|                        \\
   & =|A_{\mathrm{E,c}}-A_{\mathrm{P,c}}|
  =R_{\mathrm{pm}}.
\end{align*}
The detected erase branch has
\(|\mathcal A_{\mathrm{E,c}}|=A_{\mathrm{E,c}}
\ge R_{\mathrm{pm}}\), so
\[
  |\mathcal A_{\mathrm{E,c}}|
  +|\mathcal A_{\mathrm{comp,c}}|
  \ge 2R_{\mathrm{pm}}.
\]
Conservatively assigning all preserve-condition residual coherence to
  non-path-marked background can only decrease the inferred path-marked
  normalization and therefore does not weaken this floor.

Let \(R_{\mathrm{pm}}^{\mathrm{launch}}\) denote the rate of coherent
path-marked idlers entering the launch optics, before launch-specific mirror,
window, and propagation losses. Launched idlers are not conditioned on
transmission through the eraser polarizer, collection into the erase detector
mode, or idler-detector efficiency. Therefore
\[
  R_{\mathrm{pm}}^{\mathrm{launch}}\ge2R_{\mathrm{pm}}.
\]
This bound uses only linearity and analyzer completeness in the measured
  common mode. It does not assume a Bell-pair form, balanced horizontal and
  vertical populations, purity, or ideal extinction. It applies to the fixed
  effective projection defined by the roof-fiber unitary and eraser polarizer
  without requiring equal marker-state weights. Using a detected erase pair to
  benchmark the fringe from a coherent launched pair is conservative because
  imperfect balance of the effective erasing projection reduces the erased
  coincidence contrast, whereas launched idlers pass through no erasure optics.
  The \ac{mzi} contrast is common to both ensembles, and accidental
  coincidences carry no fringe phase and do not enter \(R_{\mathrm{pm}}\).

Applying the infinite-future and finite-path transmission factors gives
corresponding survivor rates of at least \(2R_\infty\) and
\(2R_{\mathrm{fin}}\), respectively. Applying this conservative normalization
floor to the pooled likelihood gives
\[
  \eta^{\rm launch}_{\infty,95}
  <0.078\quad\text{at 95\% confidence}.
\]
The finite-path ensemble gives \(\eta^{\rm launch}_{\mathrm{fin},95} <0.077\)
  at 95\% confidence.

We retain the coincidence-basis \(R_{\mathrm{pm}}\) as the primary
normalization because it provides the most direct and conservative connection
to the measured data.

\section{Illustrative-dataset case study for D5}
\label{app:illustrative-dataset-case-study}

This section gives a worked example using the illustrative dataset D5. It
collects the dataset-specific diagnostics and intermediate results underlying
that example.

\subsection*{Backgrounds and accidental coincidences}

The background and accidental-coincidence correction procedure is given in
Appendix~\ref{sec:experimental-setup}, under ``Dark-run backgrounds and
accidental coincidences.'' Table~\ref{tab:dark-rates} reports the background
rates and residual dark-coincidence backgrounds.
Table~\ref{tab:accidental-corrections} gives the rate budgets and
accidental-coincidence corrections. In Table~\ref{tab:accidental-corrections},
launch-condition idler and coincidence entries are N/A because the idler was
not detected.

\begin{table}[ht]
  \caption{
    Measured in-apparatus background rates \(d_X\) and derived residual
    dark-coincidence backgrounds \(b_{C_j}\) for the illustrative dataset D5.
    The dark acquisition time was \(\Delta t_{\mathrm{dark}}=60\,\mathrm{s}\) for
    every condition, and all tabulated rates are in \(\mathrm{s}^{-1}\).
  }
  \label{tab:dark-rates}
  \centering
  \begingroup
  \setlength{\tabcolsep}{3pt}
  \begin{tabular}{@{}lrrrr@{}}
    \toprule
    Quantity  & \multicolumn{2}{c}{Launch--erase run} & \multicolumn{2}{c}{Preserve--erase run}                                                           \\
    \cmidrule(lr){2-3}\cmidrule(lr){4-5}
              & Erase                                 & Launch                                  & Erase                     & Preserve                    \\
    \midrule
    $d_{S_1}$ & 564.5\ensuremath{\pm}3.1              & 559.4\ensuremath{\pm}3.1                & 577.3\ensuremath{\pm}3.1  & 577.6\ensuremath{\pm}3.1    \\
    $d_{S_2}$ & 475.0\ensuremath{\pm}2.8              & 476.6\ensuremath{\pm}2.8                & 484.8\ensuremath{\pm}2.8  & 488.5\ensuremath{\pm}2.9    \\
    $d_I$     & 3988.3\ensuremath{\pm}8.2             & 392.4\ensuremath{\pm}2.6                & 5181.7\ensuremath{\pm}9.3 & 14097.8\ensuremath{\pm}15.3 \\
    $d_{C_1}$ & 0.12\ensuremath{\pm}0.04              & 0.02\ensuremath{\pm}0.02                & 0.15\ensuremath{\pm}0.05  & 0.40\ensuremath{\pm}0.08    \\
    $d_{C_2}$ & 0.05\ensuremath{\pm}0.03              & 0.00\ensuremath{\pm}0.00                & 0.10\ensuremath{\pm}0.04  & 0.37\ensuremath{\pm}0.08    \\
    $b_{C_1}$ & 0.06\ensuremath{\pm}0.04              & 0.01\ensuremath{\pm}0.02                & 0.08\ensuremath{\pm}0.05  & 0.20\ensuremath{\pm}0.08    \\
    $b_{C_2}$ & 0.00\ensuremath{\pm}0.03              & 0.00\ensuremath{\pm}0.00                & 0.04\ensuremath{\pm}0.04  & 0.19\ensuremath{\pm}0.08    \\
    \bottomrule
  \end{tabular}
  \endgroup
\end{table}

\begin{table}[ht]
  \caption{
    Acquisition rate budgets and accidental-coincidence corrections for the
    illustrative dataset D5. All rates are in \(\mathrm{s}^{-1}\) unless otherwise indicated.
    N/A denotes launch-condition quantities for which the idler was not detected.
  }
  \label{tab:accidental-corrections}
  \centering
  \begingroup
  \setlength{\tabcolsep}{3pt}
  \begin{tabular}{@{}lrrrr@{}}
    \toprule
    Quantity                  & \multicolumn{2}{c}{Launch--erase run} & \multicolumn{2}{c}{Preserve--erase run}                     \\
    \cmidrule(lr){2-3}\cmidrule(lr){4-5}
                              & Erase                                 & Launch                                  & Erase  & Preserve \\
    \midrule
    $R_{S_1}^{\mathrm{raw}}$  & 2599.6                                & 2599.7                                  & 2602.7 & 2604.3   \\
    $R_{S_2}^{\mathrm{raw}}$  & 2074.6                                & 2074.6                                  & 2075.0 & 2078.8   \\
    $R_I^{\mathrm{raw}}$      & 9987.4                                & \text{N/A}                              & 9855.0 & 29550.1  \\
    $R_C^{\mathrm{raw}}$      & 143.47                                & \text{N/A}                              & 143.32 & 421.69   \\
    $R_{C_1}^{\mathrm{corr}}$ & 71.87                                 & \text{N/A}                              & 78.08  & 234.40   \\
    $R_{C_2}^{\mathrm{corr}}$ & 70.37                                 & \text{N/A}                              & 63.97  & 183.44   \\
    $R_C^{\mathrm{corr}}$     & 142.24                                & \text{N/A}                              & 142.06 & 417.84   \\
    $R_{\mathrm{acc}}$        & 1.17                                  & \text{N/A}                              & 1.15   & 3.46     \\
    $f_{\mathrm{acc}}$ (\%)   & 0.81                                  & \text{N/A}                              & 0.80   & 0.82     \\
    \bottomrule
  \end{tabular}
  \endgroup
\end{table}

\subsection*{Atmospheric and Galactic transmission}

The transmission procedure is given in Appendix~\ref{app:transmission-details}.
Table~\ref{tab:atm-20260308} lists the libRadtran inputs and outputs.

\begin{table}[ht]
  \caption{
    Atmospheric transmission for the illustrative dataset D5, whose launch window started on
    March 8, 2026, at 19{:}21{:}56\ \acf*{utc}
    and lasted 49\,min.
    The \acf*{metar} observation time was 18{:}53{:}00\ \acs*{utc}
    (29\,min before the launch window).
    The \acf*{goes18} file had a start time of
      19{:}21{:}18\ \acs*{utc}
    (38\,s before the launch window)
    and data-quality flags
    \acf*{tpw}=0,
    \acf*{aod}=1,
    and \acf*{cod}=134\ (invalid).
    The Ångström exponent is from the \acs*{goes18} \acs*{aod} AE1 field.
    libRadtran parameterizes \acs*{aod} as
    $\tau_a(\lambda)=\beta\,\lambda^{-\alpha}$, with $\lambda$ expressed in
    micrometers, so we derived $\beta=AOD_{550}(0.55)^{\alpha}$. AE1 denotes
    the \acs*{goes18} Ångström-exponent field, and CLR denotes clear skies.
  }
  \label{tab:atm-20260308}
  \centering
  \begin{tabular}{l c r}
    \toprule
    Quantity                     & Source        & Value           \\
    \midrule
    Filter (top-hat)             & Analysis      & 782--867\ nm    \\
    Surface pressure $P$         & \acs*{metar}  & 1013.6\ hPa     \\
    Surface temperature          & \acs*{metar}  & 22.8\ $^\circ$C \\
    Sky condition                & \acs*{metar}  & CLR             \\
    \acs*{cod}                   & \acs*{goes18} & Invalid (clear) \\
    \acs*{tpw}                   & \acs*{goes18} & 5.749\ mm       \\
    $AOD_{550}$                  & \acs*{goes18} & 0.07414         \\
    Ångström exponent $\alpha$   & \acs*{goes18} & -0.19139        \\
    Ångström coefficient $\beta$ & Derived       & 0.08313         \\
    $T_{\mathrm{atm}}$           & libRadtran    & 0.8873          \\
    \bottomrule
  \end{tabular}

  \vspace{2pt}

\end{table}

Table~\ref{tab:mw-20260308} provides the Galactic transmission details.

\begin{table}[ht]
  \caption{
    Galactic extinction for the illustrative dataset D5\ along
    the geometric zenith
    over the \acl*{ucsb} at the start of the launch window
    on March 8, 2026, at 19{:}21{:}56\ \acl*{utc}.
    The table uses \acf*{ra}, \acf*{dec}, the NASA/IPAC \acf*{irsa}, and the
    \acf*{dssii}; MW denotes Milky Way.
    \texttt{A\_SandF} denotes the extinction column
    recalibrated by Schlafly and Finkbeiner.
  }
  \label{tab:mw-20260308}
  \centering
  \begin{tabular}{l c r}
    \toprule
    Quantity         & Source            & Value                \\
    \midrule
    Zenith \acs*{ra} & Launch time       & 22{:}27{:}26.3       \\ Zenith \acs*{dec} &
    Launch time      & +34{:}17{:}09.0                          \\ Band & \acs*{irsa} & DSS-II i \\ Effective
    wavelength       & \acs*{irsa}       & 811.1\ $\mathrm{nm}$ \\ Extinction column &
    \acs*{irsa}      & \texttt{A\_SandF}                        \\ $A_\lambda$ & \acs*{irsa} & 0.171\ mag
    \\ $\tau_{\mathrm{MW}}$ & Derived & 0.157 \\ $T_{\mathrm{MW}}$ & Derived &
    0.8543                                                      \\ \bottomrule
  \end{tabular}
\end{table}

\subsection*{Fringe-fit results}

The fitting procedure is given in Sec.~\ref{supp-scan-fits}. For this dataset,
the free-period fit gave \(P=(2.077\pm0.009)\,\mathrm{V}\).
Table~\ref{tab:fits} summarizes the joint-fit parameters, including the
exit-specific baseline rates, visibility, and total fringe amplitude.
\begin{figure}[t]
  \centering
  \includegraphics{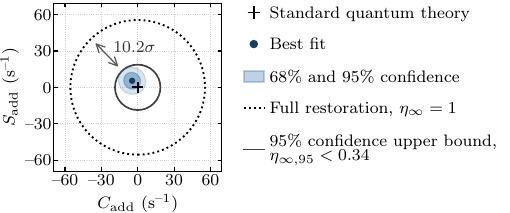}
  \caption{
    Exclusion plot for a launch-induced signal fringe in the illustrative dataset D5.
    The dot marks the best-fit launch-only quadratures
    \((\hat C_{\mathrm{add}},\hat S_{\mathrm{add}})\), with 68\% and 95\%
    shaded joint confidence regions.
    The solid circle is the 95\% confidence upper bound on an arbitrary-phase
    added fringe, \(A_{\mathrm{add},95}=18.8\,\mathrm{s}^{-1}\)
    \((\eta_{\infty,95}<0.34)\).
    The dotted circle is full restoration for the
    surviving-idler ensemble,
    \(A_{\mathrm{add}}=R_\infty\approx
    56\,\mathrm{s}^{-1}\)
    \((\eta_\infty=1)\).
    The cross marks the prediction of standard quantum mechanics,
    \(A_{\mathrm{add}}=0\).
  }
  \label{fig:null-plane}
\end{figure}
\begin{table}
  \caption{
    Summary of the paired launch--erase contrast fit for the illustrative dataset D5.
    The 95\% bound is the maximum radius of the 95\% confidence
    ellipse for \((C_{\mathrm{add}},S_{\mathrm{add}})\).
    The full-restoration exclusion \(Z_\infty\) is the raw phase-minimized
    covariance-weighted distance, retained as a conditional fit diagnostic.
    Extra digits are carried for arithmetic reproducibility, not as
    significant figures.
    Uncertainties are marginal \(1\sigma\) fit uncertainties.
  }
  \label{tab:null-bounds}
  \centering
  \begin{tabular}{@{}lr@{}}
    \toprule
    Quantity                                  & Value                               \\
    \midrule
    Paired scans                              & 9                                   \\
    Paired points                             & 99                                  \\
    \(\chi_\nu^2\) (176\ degrees of freedom)
                                              & 1.033                               \\
    \(\hat C_{\mathrm{add}}\)
                                              & \((-4.60\pm4.63)\,\mathrm{s}^{-1}\) \\
    \(\hat S_{\mathrm{add}}\)
                                              & \((5.41\pm4.60)\,\mathrm{s}^{-1}\)  \\
    \(\operatorname{Cov}(\hat C_{\mathrm{add}},\hat S_{\mathrm{add}})\)
                                              & \(-1.355\,(\mathrm{s}^{-1})^2\)     \\
    \(\hat A_{\mathrm{add}}=\sqrt{\hat C_{\mathrm{add}}^2+\hat S_{\mathrm{add}}^2}\)
                                              & \(7.101\,\mathrm{s}^{-1}\)          \\
    Two-dimensional Gaussian null \(p\) value & 0.328                               \\
    Residual sign-flip \(p\) value (10,000 draws)
                                              & 0.261                               \\
    \(A_{\mathrm{add},95}\)
                                              & \(18.746\,\mathrm{s}^{-1}\)         \\
    \(R_\infty\)                              & \(55.56\,\mathrm{s}^{-1}\)          \\
    \(\eta_{\infty,95}\)                      & 0.337                               \\
    Full-restoration exclusion, \(Z_\infty\)
                                              & \(10.2\sigma\)                      \\
    \bottomrule
  \end{tabular}
\end{table}
\subsection*{Path-marked and surviving-idler normalization}

For this dataset, Run 1 in Table~\ref{tab:fits} is the launch--erase run and
Run 2 is the preserve--erase run. The path-marked normalization uses the Run 1
erase fit and the Run 2 preserve fit. Thus,
\[
  \begin{aligned}
    R_{\mathrm{pm}}
     & =
    A_{\mathrm{E},\mathrm{c}}
    -A_{\mathrm{P},\mathrm{c}}
    \\
     & =
    \bigl[
      (81.0\pm1.0)
      -(5.4\pm1.0)
      \bigr]\,\mathrm{s}^{-1}
    \\
     & =
    (75.6\pm1.4)\,\mathrm{s}^{-1}.
  \end{aligned}
\]

Tables~\ref{tab:atm-20260308} and \ref{tab:mw-20260308} give the atmospheric
and Galactic inputs for the selected launch window.
Table~\ref{tab:dataset-transmission} combines them with the common
launch-optics and intergalactic factors. The resulting normalization is
\[
  T_\infty=0.7348,
  \qquad
  R_\infty
  =T_\infty R_{\mathrm{pm}}
  =55.56\,\mathrm{s}^{-1}.
\]
This is the reference rate reported in Table~\ref{tab:null-bounds} and used
  there to normalize the added-fringe bound from the paired launch--erase
  contrast fit. These terminal values are calculated from the unrounded
  analysis outputs.

\subsection*{Paired launch--erase contrast fit}

The fitting procedure is given in Sec.~\ref{null-bound}.
Table~\ref{tab:null-bounds} summarizes the fit results and bounds. The
launch-only quadratures are consistent with zero. Figure~\ref{fig:null-plane}
shows the fit. The 95\% amplitude bound is
\[
  A_{\mathrm{add},95}
  =18.8\,\mathrm{s}^{-1}.
\]
Normalizing by \(R_\infty\) gives
\[
  \eta_{\infty,95}
  \equiv
  \frac{A_{\mathrm{add},95}}{R_\infty}
  <
  0.34.
\]
Full restoration \((\eta_\infty=1)\) would place the added-quadrature vector
  on a circle of radius \(R_\infty\approx56\,\mathrm{s}^{-1}\), about
  \(2.5\times\) the observed launch-condition signal-singles amplitude of
  \((22\pm3)\,\mathrm{s}^{-1}\). This circle is separated from the best fit by
  a radial gap of \(48.5\,\mathrm{s}^{-1}\). On the stated nominal
  normalization, full restoration is excluded at \(>5\sigma\) for any phase.

The finite-path result follows from the same fitted amplitude bound by
replacing \(R_\infty\) with \(R_{\mathrm{fin}}\). Both results are reported in
Table~\ref{tab:dataset-analysis}.

\section{Transmission-normalization robustness}
\label{transmission-robustness}

The paired interference fits determine the launch-only quadratures
$\widehat{\mathbf q}_d$ and their covariances $\Sigma_d$ without using the
transmission model. The fitted quadratures are therefore independent of the
transmission model, and the pooled test of $\eta=0$ is unchanged by any common
rescaling of the transmissions, including the choice of idler ensemble.

Transmission enters only in normalizing a possible launch-induced fringe to a
restoration fraction. The finite-path normalization uses only the
well-constrained, finite-path components of the transmission model and gives
essentially the same pooled bound as the infinite-future normalization,
$\eta_{\mathrm{fin},95}<0.16$ versus $\eta_{\infty,95}<0.16$. More generally,
under a common rescaling $T_e^{(d)}\to fT_e^{(d)}$ for either ensemble $e$, the
upper endpoint scales as $\eta_{e,95}(f)=\eta_{e,95}(1)/f$. Within this
common-error model, full restoration would cease to be excluded only if every
transmission were overestimated by more than roughly sixfold.

The conclusion is stable over the full modeled survival-fraction range spanned
by the two ensembles, \(0.73\)--\(0.83\). Each of the six datasets individually
excludes full restoration under both normalizations
(Table~\ref{tab:dataset-analysis}). The leave-one-dataset-out values of
\(\eta_{\infty,95}\) range from \(0.154\) to \(0.168\).%

\end{document}